\documentclass[aps, prd, twocolumn, lengthcheck, superscriptaddress, 
nofootinbib]{revtex4-1}

\usepackage{amsfonts}
\usepackage{ulem}
\usepackage{amsmath}
\usepackage{amssymb}
\usepackage{xcolor}
\usepackage{graphicx}
\usepackage{slashed}
\usepackage{hyperref}
\usepackage{amsthm}
\usepackage{mathrsfs}
\usepackage{color}
\usepackage{bm}
\usepackage{ae,aecompl}	
\usepackage{soul}
\usepackage{multirow}
\usepackage{verbatim}

\def\be{\begin{eqnarray}}
\def\ee{\end{eqnarray}}

\begin{document}
\title{Compact star and compact star matter properties from a baryonic extended linear sigma model with explicit chiral symmetry breaking}

\author{Yao Ma}
\email{mayao@nju.edu.cn}
\affiliation{School of Frontier Sciences, Nanjing University, Suzhou 215163, China}

\author{Yong-Liang Ma}
\email{ylma@nju.edu.cn}
\affiliation{School of Frontier Sciences, Nanjing University, Suzhou 215163, China}
\affiliation{International Center for Theoretical Physics Asia-Pacific (ICTP-AP) , UCAS, Beijing 100190, China}

\author{Lu-Qi Zhang}
\affiliation{School of Physics, Nanjing University, Nanjing, 210093, China}
\affiliation{School of Frontier Sciences, Nanjing University, Suzhou 215163, China}

\begin{abstract}
Based on a baryonic extended linear sigma model including explicit chiral symmetry breaking effect, the structure of neutron stars with the emergence of hyperons is investigated using the relativistic mean field approximation.
It is found that, except for the lightest scalar meson $\sigma$ whose structure is not well understood so far, the vacuum mass spectra of relevant hadrons and nuclear matter properties around saturation density can be well reproduced.
Nevertheless, based on the present model and the applied relativistic mean field approach, we found that, to have a realistic mass-radius relation of neutron stars, the $\pi N$ sigma term $\sigma_{\pi N}$ that denotes the contribution of explicit symmetry breaking should deviate from its empirical values at vacuum.
Specifically, $\sigma_{\pi N}\sim -600$ MeV, rather than $(32\text{--}89) \rm \ MeV$ at vacuum.
With an appropriate choice of $\sigma_{\pi N}$ and $K(n_0)$, our framework can give a more observationally favored mass-radius relation of neutron stars with the emergence of hyperons, suggesting a possible density dependence of the low energy constants, at least within the present leading order framework with the relativistic mean field approach.
The present result provides a new perspective on the relation between microscopic explicit chiral symmetry breaking in dense matter and macroscopic structure of compact stars and calls for more systematic treatments beyond leading order relativistic mean field calculation.

\end{abstract}

\maketitle

\allowdisplaybreaks

\section{Introduction}

The properties of dense nuclear matter (NM) and structures of neutron stars (NSs) have become one of the most active and challenging topics, especially since the beginning of multi-messenger era of NS observations~\cite{LIGOScientific:2017ync}.
A NS is a compact object mainly composed of nucleons admixed with other hadronic degrees of freedom. Many studies have been performed to understand its equation of state (EOS)~\cite{Akmal:1998cf,Lattimer:2000nx,Lattimer:2004pg,Steiner:2010fz,Hebeler:2013nza,Baym:2017whm,Ma:2019ery,Brandes:2023bob}.
The mass-radius (M-R) relation of a NS is determined by the balance between the gravitational force and the pressure of the matter inside via Einstein's equation, practically obtained by solving the Tolman-Oppenheimer-Volkoff (TOV) equation~\cite{Tolman:1939jz,Oppenheimer:1939ne}.
The TOV equation is a self-consistent equation of matter distribution from the dense core to the surface, so that the M-R relation depends on a wide density region of the EOS, say, up to \((5\text{--}10) n_0\) with $n_0 \approx 0.16$~fm$^{-3}$ being the saturation density~\cite{Shapiro:1983du}.

It was predicted that, in the core of massive stars, the strange freedom may be present due to $\beta$ equilibrium of weak interactions~\cite{Witten:1984rs}.
A lot of work has been done to investigate the possible existence of strange quarks in the core of NSs~\cite{Gal:2016boi,Weissenborn:2011ut,Weissenborn:2011kb,Chatterjee:2015pua,Weber:2004kj,Weissenborn:2011qu,Ozel:2010bz}.
When considering strange degrees of freedom at the hadron level, the lightest hyperon $\Lambda$ is the one most likely to be produced and  remain stable.
Hyperons can emerge at the densities \(\sim 2n_0\)~\cite{Schaffner-Bielich:2008zws} which is the core density for NSs with mass $\gtrsim  1.4~M_\odot$.
The earliest work on the hyperon in NSs can be traced back to 1960~\cite{ambartsumyan1960degenerate}, but there are still many puzzles about the influence of hyperons on NS structures, for example, the EOS is softened due to the emergence of hyperons and, as a result, the upper bound of the yield NS mass is reduced to below the canonical one~\cite{Schulze:1995jx,Katayama:2014gca,Zdunik:2012dj,Oertel:2014qza}.
The observation of NSs with masses  $\sim 2M_\odot$, in particular PSR J1614-2230~\cite{Demorest:2010bx} and PSR J0348+0432~\cite{Antoniadis:2013pzd},  critically sharpened this puzzle, as most hyperon-included EOS models fail to support such high masses~\cite{Bombaci:2016xzl,Tolos:2020aln}.
So far, there have been many attempts to cope with this problem, e.g., including the repulsive interactions between hyperons through vector meson exchanges~\cite{Glendenning:1982nc,Hofmann:2000vz,bednarek2012hyperons,Lim:2014sra}, three-body interactions~\cite{takatsuka2002necessity,Lonardoni:2014bwa}, and proposing a phase transition to quark matter~\cite{alford2007quark,Klahn:2013kga,Masuda:2012kf}. However, most of these works lack the full consideration of a rigorous relation to QCD.

More recently, additional strategies have been explored. Microscopically, SU(3) chiral effective field theory for the $\Lambda N$ interaction found that the $\Lambda$ single-particle potential  becomes strongly repulsive at densities of $(2\text{--}3)n_0$, which could push back the hyperon onset~\cite{Haidenbauer:2016vfq}. In addition, the hyperonic three-body forces are expected to contribute further repulsion~\cite{Vidana:2010ip}. Quantum many-body effects from the strong baryon-meson coupling were found to sufficiently stiffen the hyperon-star EOS~\cite{Zhu:2026mhd}, while chiral symmetry restoration in SU(3) parity doublet models was argued to offer a natural mechanism for suppressing hyperon emergence without \emph{ad hoc} repulsion~\cite{Gao:2026mla}. On the experimental side, the upcoming K-long Facility~\cite{Zachariou:2024omo} and hypernuclei-beam heavy-ion experiments~\cite{Yong:2025dpe} aim to constrain the hyperon potential across a wide density range. Furthermore, an artificial-intelligence-assisted nuclear emulsion analysis has recently achieved the first uniquely identified double-$\Lambda$ hypernucleus~\cite{He:2025ndk}.

Recently, we investigated the properties of NM based on an extended linear sigma model with baryons (bELSM)~\cite{Ma:2023eoz}.
The Lagrangian was constructed with three-flavor chiral symmetry in order to mix the two- and four-quark configurations to reproduce the mass spectra of the light scalar mesons~\cite{Fariborz:2005gm,Fariborz:2007ai,Fariborz:2009cq}.
Since bELSM is constructed following the chiral symmetry pattern of QCD with consistent power counting laws~\cite{Ma:2023eoz,Fariborz:2005gm,Fariborz:2007ai,Fariborz:2009cq,Parganlija:2010fz}, it is rather interesting to see how the EOS behaves in the presence of hyperons in this framework and whether there is any new insight into the relation between microscopic symmetries and NS structures.

Regarding the mass splitting of hadrons stemming from the explicit breaking of QCD flavor symmetry, which is needed to simulate the $\beta$ equilibrium, we should go beyond the Lagrangian introduced in Ref.~\cite{Ma:2023eoz} and include the explicit chiral symmetry breaking terms~\cite{Lenaghan:2000ey} (termed as bELSM-\(\xi\)). In this work, the Lagrangian is constructed at the lowest order of the explicit chiral breaking by introducing a constant scalar background field that is matched to the quark mass terms in the QCD Lagrangian~\cite{Scherer:2002tk}. With an appropriate choice of parameters, the Gell-Mann--Okubo mass formula of baryons~\cite{Gell-Mann:1964ewy,Okubo:1961jc} is reproduced, which therefore links the phenomenon related to $\beta$ equilibrium with the microscopic symmetries. The free parameters of the model are determined by the mass spectra of hadrons at vacuum and the NM properties around saturation density $n_0$  are calculated with the help of relativistic mean field (RMF) approximation~\cite{Serot:1984ey,Serot:1997xg}. After that, the impact of hyperons on the M-R relation of NSs is studied with the TOV equation~\cite{Tolman:1939jz,Oppenheimer:1939ne}.

It is found that the vacuum mass spectra of relevant hadrons and NM properties can be well reproduced except for the lightest scalar meson $\sigma$ whose structure is still under debate.
However, when one wants to have a reasonable M-R relation within the constraints of MSP J0740+6620~\cite{NANOGrav:2019jur} and GW170817~\cite{LIGOScientific:2018cki}, either the incompressibility at $n_0$, \(K(n_0)\), or \(\sigma_{\pi N}\) should deviate from their empirical values and are found to be \(K(n_0)\sim 500\) or \(\sigma_{\pi N}\sim -600 \rm \ MeV\) in contrast to their empirical values \(K(n_0) \approx (200\text{--}260)\)~\cite{Sedrakian:2022ata} and \(\sigma_{\pi N}\sim (32\text{--}89) \ \rm MeV\) at tree level estimation without isospin breaking~\cite{Bernard:1995dp,Meissner:2005ba}.
The necessary value \(K(n_0)\sim 500\rm \ MeV\) was already found in the previous bELSM studies~\cite{Ma:2023eoz,Ma:2025llw}, and the present three-flavor analysis including the explicit symmetry breaking confirms this tension.
It should be emphasized that these observations are drawn at the leading order (LO) calculation with the RMF approach, and the tension may be partially due to the present approach, e.g., the truncation of the model and the RMF approximation. This observation may also suggest the possible density dependence of the low energy constants, and calls for a more systematic investigation beyond the LO.

Furthermore, the impact of hyperons on the M-R relation is found to be sensitive to the density where the hyperons emerge.
This density is determined by the competition between the couplings of vector mesons to the nucleon and hyperon, \(g_{\omega NN}\) and \(g_{\omega \Lambda\Lambda}\).
At LO in bELSM-\(\xi\) \(g_{\sigma NN}= g_{\sigma \Lambda\Lambda}\), to yield the empirical estimation of the hyperon potential \(U_{\Lambda}\sim -28\ \mathrm{MeV}\)~\cite{Millener:1988hp,Glendenning:1991es}, we therefore obtained \(g_{\omega NN}\simeq g_{\omega \Lambda\Lambda}\), consistent with flavor symmetry. As a result, the hyperon emerges at a higher density than that in Walecka-type models, which makes the maximum mass of NSs to not decrease so much compared to the two-flavor case.

The rest of this paper is organized as follows:
In Sec.~\ref{sec:model}, we introduce the Lagrangian with the explicit symmetry breaking at the LO and the mass splitting of baryons and mesons.
In Sec.~\ref{sec:sigpiN}, we study the effect of the $\pi$-$N$ sigma term on the NS structures. In Sec.~\ref{sec:NS}, we analyze the EOS at densities relevant to the core of NSs. The effects of hyperons on NS structure are analyzed in Sec.~\ref{sec:LamNS}.
Finally, summary and outlook are given in Sec.~\ref{sec:conclusion}. The details of the Lagrangian construction, the EOS calculation, and the Waleck-type model compared are given in Appendixes.

\section{bELSM with explicit symmetry breaking under RMF approximation}

\label{sec:model}

The details of the bELSM model construction without including the quark masses were discussed in Refs.~\cite{Ma:2023eoz,Ma:2025llw}.
When it comes to the three-flavor case, the main concern is the possible existence of strange degrees of freedom due to the $\beta$ equilibrium of weak interaction, where the mass splitting plays an important role. Therefore, it is necessary to extend the bELSM to incorporate the quark mass effect from the microscopic point of view.

The QCD Lagrangian is invariant under the transformation of the flavor group $\rm U(3)_L\otimes \rm U(3)_R$ if one neglects the quark masses.
The quark mass slightly but explicitly breaks chiral symmetry, and can be regarded as a constant scalar background field $\xi$ coupled to the quark fields~\cite{Scherer:2002tk}.

In this work, we regard this scalar background \(\xi\) as a perturbation field and only consider the terms at the LO of $\xi$.  The RMF Lagrangian can be written as follows: 
\be
\mathcal{L}=\mathcal{L}_{\rm M}+\mathcal{L}_{\rm V}+\mathcal{L}_{\rm B}\ ,
\label{eq:LagLO}
\ee
where 
\begin{subequations}
\be
\label{eq:LM}
\mathcal{L}_{\rm M} & = & c_2 \operatorname{Tr}S'^2-c_4 \operatorname{Tr}S'^4 \nonumber\\
& &{} -b_1 \operatorname{Tr}\left({\xi}S'^3\right)-2G \operatorname{Tr}\left({\xi}S'\right)\ ,\\
\label{eq:LV}
\mathcal{L}_{\rm V} & = & \tilde{h}_2 \operatorname{Tr}\left(V^2S'^2\right)+a_1 \epsilon^{i j k} \epsilon^{l m n}(V)_{i l}(V)_{j m}\left(S'^2\right)_{k n} \nonumber\\
& &{} +b_2 \operatorname{Tr}\left(V^2 {\xi} S'\right) + b_3 \epsilon^{i j k} \epsilon^{l m n}(V)_{i l}(V)_{j m}\left({\xi}S'\right)_{k n} \nonumber\\
& &{} +\tilde{g}_3 \operatorname{Tr}V^4+\tilde{a}_2 \epsilon^{i j k} \epsilon^{l m n}(V)_{i l}(V)_{j m}(V^2)_{k n}\ , \\
\label{eq:LB}
\mathcal{L}_{\rm B} & = & \operatorname{Tr}\left(\bar{B} i \gamma_\mu \partial^\mu B\right) +c \operatorname{Tr}\left[\bar{B} \gamma_0 V B\right]+c^{\prime} \operatorname{Tr}\left[\bar{B} \gamma_0 B V\right] \nonumber\\
& &{} +h \epsilon^{i j k} \epsilon^{l m n}(\bar{B})_{i l}\left(\gamma_0 V \right)_{j m}(B)_{k n} -g \operatorname{Tr}[\bar{B} S' B] \nonumber\\
& &{} -e \epsilon^{i j k} \epsilon^{l m n}(\bar{B})_{i l}(S')_{j m}(B)_{k n}  -b_4 \operatorname{Tr}[\bar{B} {\xi} B]\nonumber\\
& &{} -b_5 \epsilon^{i j k} \epsilon^{l m n}(\bar{B})_{i l}({\xi})_{j m}(B)_{k n}\ .
\ee
\end{subequations}
The full Lagrangian without RMF simplification is given in Appendix~\ref{app:1st-L}.
After taking the RMF approximation, the relevant meson fields take the forms
\be
S' & = & \left(\begin{array}{ccc}
    \frac{\sigma^\prime}{\sqrt{3}} + \frac{f_0^\prime}{\sqrt{6}} + \frac{a_0^\prime}{\sqrt{2}} & 0 & 0 \\
    0 & \frac{\sigma^\prime}{\sqrt{3}} + \frac{f_0^\prime}{\sqrt{6}} - \frac{a_0^\prime}{\sqrt{2}} & 0 \\
    0 & 0 & \frac{\sigma^\prime}{\sqrt{3}} - \frac{2f_0^\prime}{\sqrt{6}} 
    \end{array}\right), \nonumber\\
\xi & = & \left(\begin{array}{ccc}
    \frac{\xi_0}{\sqrt{3}} + \frac{\xi_8}{\sqrt{6}} + \frac{\xi_3}{\sqrt{2}} & 0 & 0 \\
    0 & \frac{\xi_0}{\sqrt{3}} + \frac{\xi_8}{\sqrt{6}} - \frac{\xi_3}{\sqrt{2}} & 0 \\
    0 & 0 & \frac{\xi_0}{\sqrt{3}} - \frac{2\xi_8}{\sqrt{6}} 
    \end{array}\right), \nonumber\\
V & = & \frac{1}{2}\left(\begin{array}{ccc}
    \rho+\omega & 0 & 0 \\
    0 & -\rho+\omega & 0 \\
    0 & 0 & \sqrt{2} \phi
    \end{array}\right)\ ,
\ee
and the baryon fields are arranged as
\begin{equation}
    B=\left(\begin{array}{ccc}
    \frac{\Lambda}{\sqrt{6}}+\frac{\Sigma^0}{\sqrt{2}} & \Sigma^{+} & p \\
    \Sigma^{-} & \frac{\Lambda}{\sqrt{6}}-\frac{\Sigma^0}{\sqrt{2}} & n \\
    \Xi^{-} & \Xi^0 & -\frac{2 \Lambda}{\sqrt{6}}
    \end{array}\right)\ .
\end{equation}

The explicit chiral breaking results in the following vacuum condensation patterns:
\begin{equation}
    \label{eq:vev}
    \begin{aligned}
        S'=\frac{1}{\sqrt{2}}\left[\left(\alpha_3+a_0\right)\lambda_3+\left(\alpha_8+f_0\right)\lambda_8\right]+\frac{1}{\sqrt{3}}I\left(\alpha_0+\sigma\right)\ ,
    \end{aligned}
\end{equation}
where \(\alpha_i\) refers to the vacuum expectation value of the scalar meson fields, and \(a_0\), \(f_0\), and \(\sigma\) are the physical perturbation fields.

From Lagrangian \eqref{eq:LagLO}, one can obtain the following relations of coupling constants:
\be
g_{\rho NN} & = &{} -\frac{1}{2}(c+h)\ ,\nonumber\\
g_{\omega NN} & = &{} -\frac{1}{2}(c-h)\ ,\nonumber\\
g_{\omega \Lambda\Lambda} & = &{} -\frac{1}{6}(c+c'-4h)\ ,\nonumber\\
g_{\sigma NN} & = & g_{\sigma\Lambda\Lambda}=-\frac{1}{\sqrt{3}}(g-e)\ ,
\ee
and reproduce the Gell-Mann--Okubo mass formula of baryon octets,
\begin{equation}
    \label{eq:GMO}
    \begin{aligned}
        m_{\Sigma^+}&=\frac{1}{2}(-2m_n+3m_\Lambda-2m_{\Xi^-}+3m_{\Sigma^0})\ ,\\
        m_{\Sigma^-}&=\frac{1}{2}(2m_n-3m_\Lambda+2m_{\Xi^-}+m_{\Sigma^0})\ ,\\
        m_{\Xi^0}&=-m_n-m_p+3m_\Lambda-m_{\Xi^-}+m_{\Sigma^0}\ .\\
    \end{aligned}
\end{equation}
However, the mass relation of scalar mesons
\begin{equation}
    m_{a_0}^2+m_{f_0}^2=2m_{\sigma}^2\ .
\end{equation}
This yields \(m_{\sigma}\sim 1~\rm GeV\), deviating from the empirical value~\cite{ParticleDataGroup:2024cfk}. This problem can be resolved by introducing both two- and four-quark configurations of scalar mesons~\cite{Ma:2023eoz,Fariborz:2005gm,Fariborz:2007ai,Fariborz:2009cq}. We leave the detailed discussion to future work. The EOS calculation based on RMF approximation with $\beta$ equilibrium is listed in Appendix~\ref{app:eos}.

\section{Effect of \(\sigma_{\pi N}\) on hadron spectrum and nuclear matter properties}
\label{sec:sigpiN}

In principle, the free parameters in the model can be determined by the mass spectra of relevant hadrons and the properties of NM at saturation density.
In the analysis, we also find that the pion-nucleon sigma term is crucial. It is defined as~\cite{Scherer:2009bt}
\be
\sigma_{\pi N} & = & \frac{1}{2m_N} \langle N(p,s)|\mathcal{H}_{sb}(0)|N(p,s)\rangle \nonumber\\
& = & M^2 \frac{\partial m_N}{\partial M^2}
\ee
with the help of the Hellmann-Feynman theorem applied to the nucleon mass. Here, $\mathcal{H}_{sb}$ is the QCD Lagrangian including quark masses, $M^2 = 2\hat{m}B$ with $\hat{m} = (m_u+m_d)/2$, and $B$ is proportional to scalar quark condensate.
By using Lagrangian~\eqref{eq:LagLO}, the pion-nucleon sigma term can be explicitly written as
\begin{equation}
    \sigma_{\pi N}=\frac{-3\sqrt{2}\alpha_8 g_{\sigma NN}+\sqrt{3}(b_4-b_5)(2\xi_0+\sqrt{2}\xi_8)}{6}\ ,
\end{equation}
which denotes the contribution of explicit chiral symmetry breaking to the baryon nucleon mass in isospin limit.
We will see that this term is significant for describing the compact star matter properties and is helpful to understand the properties of EOS. 

During the numerical analysis, we found and will show later that, the parameter set that yields \(\sigma_{\pi N}\sim 75 \rm \ MeV\) in iso-limit that satisfies the empirical estimations at vacuum~\cite{Bernard:1995dp,Meissner:2005ba}, has difficulty in describing the high density region behavior of the EOS under RMF approximation, though it yields the reasonable masses of hadrons and properties of NM at saturation density. Therefore, we choose several values of \(\sigma_{\pi N}\), which yield the mass spectra of hadrons and NM properties at saturation density, to show its effect on the EOS in the high density region.

    The values of the free parameters, fitted by the hadron spectra [except \(\Sigma^{\pm}\) and \(\Xi^0\) due to Eq.~\eqref{eq:GMO}] and NM properties at saturation density listed in Tables~\ref{tab:mass} and \ref{tab:nm}, are given in Table~\ref{tab:para}, following the \(\chi^2\) minimization method with the central values and error bars in the two tables.

It is also found in the numerical analysis that the results are not sensitive to $G$, $\alpha_8$ and $b_5$, so these three parameters are fixed as constants across the different parameter sets.
In Table~\ref{tab:para},  \(x_{\omega}=g_{\omega\Lambda\Lambda}/g_{\omega NN}\) accounts for the competition between repulsive potential of hyperons and nucleons with  \(g_{\omega\Lambda\Lambda}\) fixed by setting the hyperon potential \(U_{\Lambda}=g_{\omega \Lambda\Lambda}\langle\omega\rangle_{n_0}-g_{\sigma \Lambda\Lambda}\langle\sigma\rangle_{n_0}=-28\rm \ MeV\)~\cite{Millener:1988hp,Glendenning:1991es}.
From Table~\ref{tab:mass}, hadron masses~\cite{ParticleDataGroup:2024cfk} can be well reproduced from the LO Lagrangian with explicit symmetry breaking, with small differences among different parameter sets from the numerical fitting. The exception is the lightest scalar meson \(\sigma\), whose disagreement indicates that additional factors, such as tetraquark configurations~\cite{Fariborz:2005gm,Fariborz:2007ai,Fariborz:2009cq,Ma:2023eoz}, need to be considered.
The mass splittings among different baryons, such as the \(\Lambda\)--\(\Sigma\)--\(\Xi\) splitting, arise from explicit flavor SU(3) breaking by quark masses. This breaking is encoded in the \(\xi\) field of the Lagrangian, directly reflecting the QCD symmetry pattern.
The nucleon mass \(m_N\) can be divided into \(m\) and \(\sigma_{\pi N}\) accounting for the mass origins from spontaneous and explicit chiral symmetry breaking, respectively, as 
\be
m_N=m+\sigma_{\pi N}.
\ee
One can read out their magnitudes from the parameter list in Table~\ref{tab:para} and hadron spectra in Table~\ref{tab:mass}.
Here, we take the isospin limit.

\begin{table}[htpb]
    \caption{
        The choices of free parameters.
       \(\sigma_{\pi N}\)-\(75^{+}\) refers to the parameter set with \(\sigma_{\pi N}\sim +75 \rm \ MeV\), and the other sets follow the same convention.
        }
    \label{tab:para}
        \begin{tabular}{@{}cccccc}
    \hline
    \hline
    &\(\sigma_{\pi N}\)-\(75^{+}\)&\(\sigma_{\pi N}\)-\(100^{-}\)&\(\sigma_{\pi N}\)-\(400^{-}\)&\(\sigma_{\pi N}\)-\(600^{-}\)\\
    \hline
    \(c_2(\times10^5\rm \ MeV^2)\)&$2.380$&$2.257$&$2.161$&$2.172$\\
    \hline
    \(G(\times10^4\rm \ MeV^2)\)&$2.623$&$2.623$&$2.623$&$2.623$\\
    \hline
    \(\alpha_3(\times10^{-2}\rm \ MeV)\)&$-8.663$&$-7.602$&$-6.177$&$-5.376$\\
    \hline
    \(\alpha_8(\rm \ MeV)\)&$15.53$&$15.53$&$15.53$&$15.53$\\
    \hline
    \(\xi_0(\rm \ MeV)\)&$-60.26$&$-26.10$&$53.13$&$132.5$\\
    \hline
    \(\tilde{h}_2\)&$179.5$&$245.7$&$356.0$&$438.7$\\
    \hline
    \(\tilde{g}_3\)&$656.0$&$442.4$&$215.5$&$234.7$\\
    \hline
    \(a_1\)&$2277$&$737.6$&$-129.8$&$-370.8$\\
    \hline
    \(b_1\)&$-28.92$&$-25.08$&$-20.55$&$-18.11$\\
    \hline
    \(b_2\)&$-164.5$&$-145.1$&$-120.4$&$-106.0$\\
    \hline
    \(b_3\)&$832.7$&$278.2$&$-58.90$&$-209$\\
    \hline
    \(b_4\)&$-8.098$&$-7.098$&$-5.861$&$-5.129$\\
    \hline
    \(b_5\)&$-0.2392$&$-0.2392$&$-0.2391$&$-0.2391$\\
    \hline
    \(g\)&$28.09$&$31.07$&$36.11$&$29.81$\\
    \hline
    \(g_{\sigma NN}\)&$-9.350$&$-11.29$&$-14.51$&$-16.93$\\
    \hline
    \(g_{\omega NN}\)&$1.795$&$4.753$&$8.572$&$10.94$\\
    \hline
    \(g_{\rho NN}\)&$5.441$&$4.661$&$3.204$&$2.510$\\
    \hline
    \(x_{\omega}\)&$2.211$&$1.364$&$1.178$&$1.149$\\
    \hline
    \hline
    \end{tabular}
\end{table}

\begin{table*}[htpb]
    \caption{
        Mass spectra of relevant hadrons in the unit of \(\rm MeV\) from the parameter sets in Table~\ref{tab:para}. The empirical values of masses are the real part of masses given in Ref.~\cite{ParticleDataGroup:2024cfk}.
        The constraints on the \(\sigma_{\pi N}\) are given in Refs.~\cite{Bernard:1995dp,Meissner:2005ba}, and the empirical values of \(m\) are given by estimation of \(m_N=(m_n+m_p)/2\) and  \(\sigma_{\pi N}\).
        }
    \label{tab:mass}
        \begin{tabular}{@{}ccccccc}
    \hline
    \hline
    &Empirical&\(\sigma_{\pi N}\)-\(75^{+}\)&\(\sigma_{\pi N}\)-\(100^{-}\)&\(\sigma_{\pi N}\)-\(400^{-}\)&\(\sigma_{\pi N}\)-\(600^{-}\)\\
    \hline
    \(m_n\)&$939.565\pm (5\times10^{-7})$&$939.565$&$939.565$&$939.565$&$939.565$\\
    \hline
    \(m_p\)&$938.272\pm (2.9\times10^{-7})$&$938.272$&$938.272$&$938.272$&$938.272$\\
    \hline
    \(m_{\Lambda}\)&$1115.68\pm(0.006)$&$1115.59$&$1115.54$&$1115.57$&$1115.58$\\
    \hline
    \(m_{\Sigma^+}\)&$1189.37\pm0.06$&$1192.30$&$1191.21$&$1191.74$&$1191.79$\\
    \hline
    \(m_{\Sigma^0}\)&$1192.64\pm0.02$&$1192.17$&$1192.18$&$1192.84$&$1191.98$\\
    \hline
    \(m_{\Sigma^{-}}\)&$1197.45\pm0.03$&$1192.03$&$1191.15$&$1191.95$&$1192.16$\\
    \hline
    \(m_{\Xi^{0}}\)&$1314.82\pm0.21$&$1331.34$&$1330.65$&$1330.90$&$1330.90$\\
    \hline
    \(m_{\Xi^{-}}\)&$1321.70\pm0.09$&$1329.77$&$1329.30$&$1329.82$&$1329.98$\\
    \hline
    \(m_{a_0}\)&$995.000\pm25.000$&$980.132$&$979.998$&$980.024$&$980.040$\\
    \hline
    \(m_{f_0}\)&$995.000\pm15.000$&$990.119$&$989.962$&$989.971$&$989.974$\\
    \hline
    \(m_{\sigma}\)&$475.000\pm75.000$&$985.183$&$984.993$&$986.010$&$985.019$\\
    \hline
    \(m_{\rho}\)&$773.000\pm2.000$&$775.261$&$775.260$&$775.260$&$775.260$\\
    \hline
    \(m_{\omega}\)&$782.660\pm0.13$&$782.659$&$782.660$&$782.660$&$782.660$\\
    \hline
    \(m_{\phi}\)&$1019.46\pm0.02$&$1019.46$&$1019.46$&$1019.46$&$1019.46$\\
    \hline
    \(\sigma_{\pi N}\)&$60.5000\pm28.5000$&$75.0541$&$-104.048$&$-402.087$&$-625.663$\\
    \hline
    \(m\)&$878.419\pm29.1465$&$863.946$&$1043.05$&$1341.09$&$1564.66$\\
    \hline
    \hline
    \end{tabular}
\end{table*}

We next consider the NM properties by using the parameters estimated in Table~\ref{tab:para}. Explicitly, we study the coefficients of Taylor expansions of energy per nucleon \(E=\mathcal{E}/n \), with respect to the density \(n=n_n+n_p\) and asymmetry \(\alpha = (n_n-n_p)/n\),
\be
E(n,\alpha) & \simeq & E(n_0,0) + E_{\rm sym}(n_0)\alpha^2 + L(n_0)\alpha^2\chi \nonumber\\
& &{} +\frac{K(n_0)}{2!} \chi^2 + \cdots\nonumber\\
& = & E_0 + \left.\frac{\alpha^2}{2}\frac{\partial^2 E(n_0, \alpha)}{\partial \alpha^2}\right|_{\alpha=0} \nonumber\\
& &{} +\alpha^2\chi\left.\left(3 n \frac{\partial E_{\mathrm{sym}}(n)}{\partial n}\right)\right|_{n=n_0} \nonumber\\
& &{} +\frac{\chi^2}{2!}\left.\left(9 n^2 \frac{\partial^2 E_0(n)}{\partial n^2}\right)\right|_{n=n_0} + \cdots \ ,
\ee
where \(\chi \equiv\left(n-n_0\right) / 3 n_0\) and $\cdots$ stands for higher orders of the expansion with respect to $\chi$ and $\alpha$. \(L(n_0)\) and \(K(n_0)\) refer to the symmetry energy slope, incompressibility of NM at \(n_0\), respectively.

After taking the standard RMF approach~\cite{Ma:2023eoz}, one can obtain the EOS in the $\beta$ equilibrium, as given in Appendix~\ref{app:eos}, and consequently the properties of NM around saturation density are given in Table~\ref{tab:nm}. From Table~\ref{tab:nm}, one can conclude that our parameter sets can well reproduce the NM properties around saturation density and $\sigma_{\pi N}$ affects the symmetry energy slope more significantly.

\begin{table}[htpb]
    \caption{
        The properties of NM at saturation density.
        The empirical values are from Refs.~\cite{Sedrakian:2022ata,Lattimer:2012xj,Dutra:2012mb}. $n_0$ is in units of fm$^{-3}$ and the others are in units of MeV.
        }
    \label{tab:nm}
\footnotesize
        \begin{tabular}{@{}ccccccc}
    \hline
    \hline
    &Empirical&\(\sigma_{\pi N}\)-\(75^{+}\)&\(\sigma_{\pi N}\)-\(100^{-}\)&\(\sigma_{\pi N}\)-\(400^{-}\)&\(\sigma_{\pi N}\)-\(600^{-}\)\\
    \hline
    \(n_0\)&$0.1550\pm0.0500$&$0.1592$&$0.1592$&$0.1592$&$0.1592$\\
    \hline
    \(E_0\)&$-15.00\pm1.00$&$-16.00$&$-16.00$&$-16.02$&$-16.00$\\
    \hline
    \(K\)&$230.0\pm30.0$&$240.8$&$241.3$&$241.6$&$239.2$\\
    \hline
    \(E_{\rm sym}\)&$30.90\pm1.90$&$30.00$&$30.00$&$30.01$&$30.00$\\
    \hline
    \(L\)&$52.50\pm17.50$&$54.91$&$52.18$&$70.94$&$76.74$\\
    \hline
    \hline
    \end{tabular}
\end{table}

\section{Equation of state and neutron star}
\label{sec:NS}

After discussing the parameter estimation and NM properties around saturation density, we are now ready to study the NS properties by extrapolating the EOS obtained above to compact star density.

\begin{figure}[htpb]
    \centering
        \includegraphics[width=0.45\textwidth]{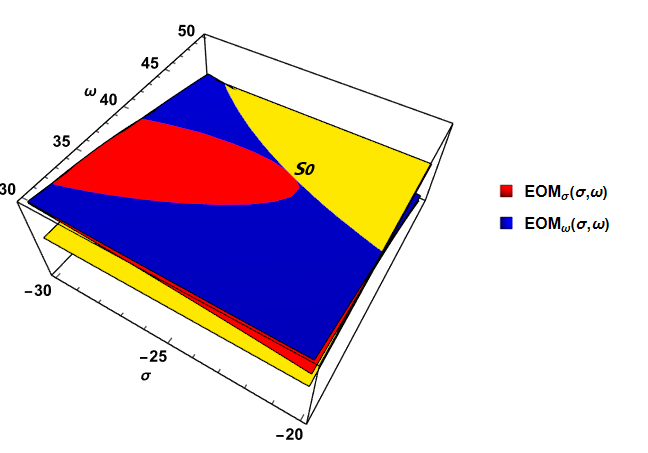}
        \includegraphics[width=0.45\textwidth]{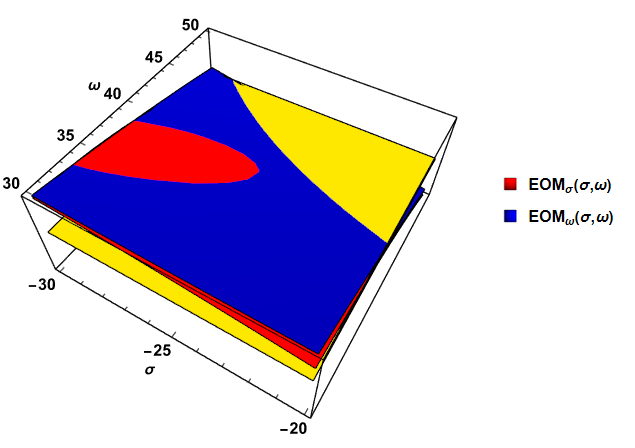}
    \caption{
        The EOM of $\sigma$ and \(\omega\) defined in Eqs.~\eqref{eq:eom-omega} and ~\eqref{eq:eom-sigma}, for pure neutron matter as functions in \(\sigma-\omega\) space. 
        The red plane is the EOM of \(\sigma\), the blue plane is the EOM of \(\omega\), and the yellow one is the zero plane.
        The intersection of the three planes is the solution of the EOM.
    }
    \label{fig:sol-plane}
\end{figure}

However, during the extrapolation, we find that when using \(\sigma_{\pi N}\)-\(75^+\), which satisfies all physical constraints listed in Tables~\ref{tab:mass} and~\ref{tab:nm}, the real number solution of the RMF equation of motion (EOM) vanishes after certain density. In Fig.~\ref{fig:sol-plane}, we illustrate the solutions of EOM of $\omega$ and $\sigma$ fields at densities $n=4.015~n_0$ (upper panel) and $n=4.050~n_0$ (lower panel). It is found that, at densities below \(4.015n_0\), the EOM have two real number solutions, one for the physical solution developed from the vacuum solution and the other is the unphysical solution that is not zero at vacuum, but the number of solutions reduces to 1 around \(4.015n_0\). However, after \(4.015n_0\) no real number solution can be found, as shown in the lower panel for \(n = 4.050~n_0\). This observation indicates that the EOM are not applicable for neutron matter above such density with this parameter set because of the fact that the mean field $\sigma$ should be real by definition.

\begin{figure}[htpb]
    \includegraphics[width=0.5\textwidth]{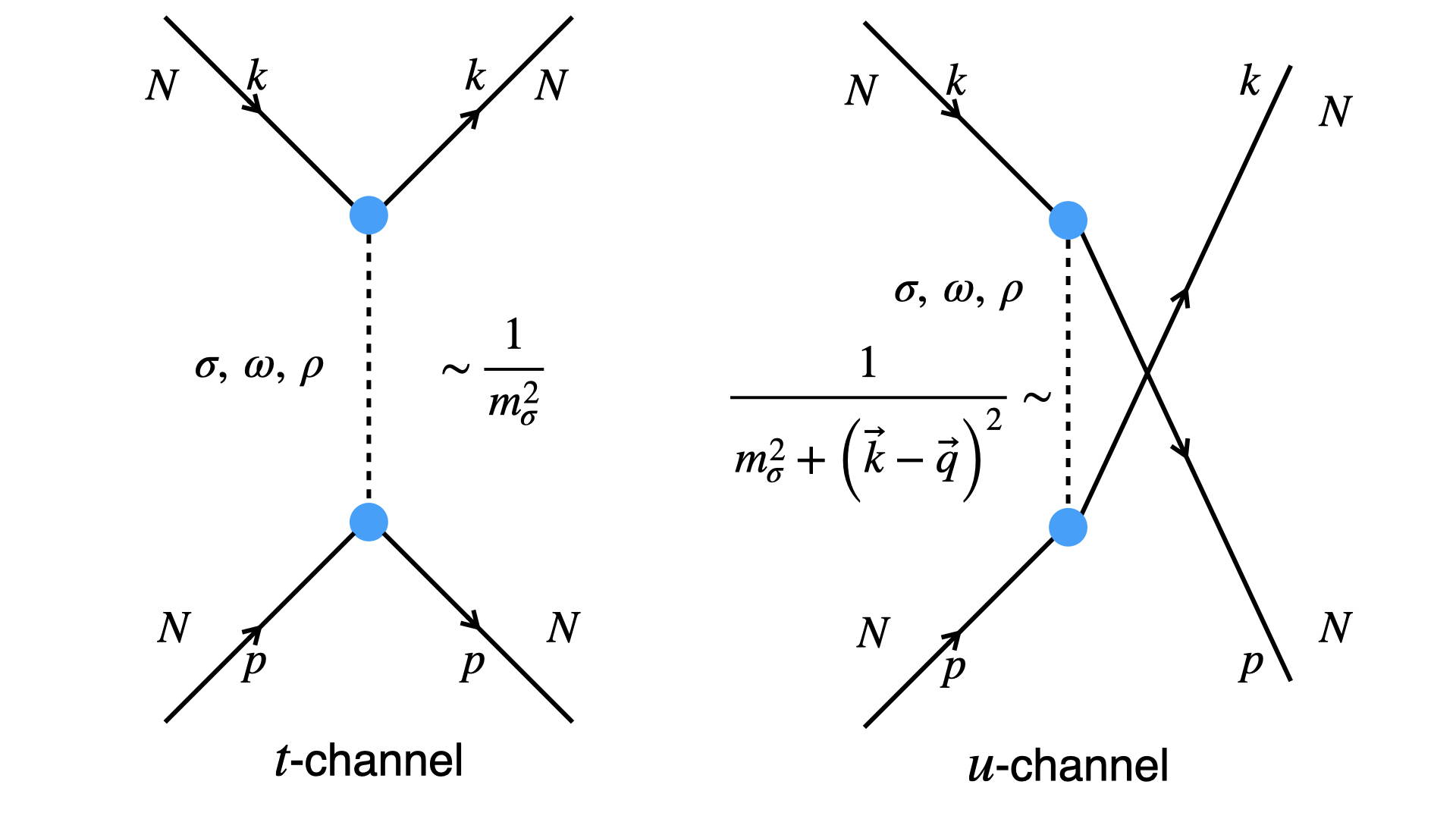}
    \caption{
        The \(t\)- and \(u\)-channel interactions of nucleons at densities illustrated in the one-boson-exchange picture.
        The transferred momentum in three-momentum space can be obtained by Fermi-Dirac distribution.
        The incoming and outgoing nucleons have the same momentum due to Fermi-Dirac statistics at zero temperature.
    }
    \label{fig:N-t-u}
\end{figure}

The reason for the invalidity of the extrapolation to higher density might be due to the limitation of the RMF approximation (see, e.g., Ref.~\cite{Xiong:2025jxq}). For example, there is no cross symmetry for baryons when treating meson fields as classical ones, that is, the \(u\)-channel interactions are not included rigorously but mimicked by the \(t\)-channel interactions with modification of the corresponding couplings, see Fig.~\ref{fig:N-t-u}. In the $u$ channel, the maximum of \(|\vec{k}-\vec{p}|\) is \(|2\vec{k}_F|\), where \(\vec{k}_F\) is the Fermi momentum of nucleons.
It can be seen that \(u\)-channel interactions can be expanded as Taylor series of \(|\vec{k}-\vec{p}|^2/m_{\sigma}^2\) at low densities.
At the LO, the \(u\)-channel interactions are mocked by the \(t\)-channel ones by adjusting the one-boson-exchange couplings, e.g., \(g_{\sigma NN}\), \(g_{\omega NN}\), and \(g_{\rho NN}\).
The convergence region of the Taylor series can be estimated as \(|2\vec{k}_F|^2<m_{\sigma}^2\), resulting in a maximum Fermi momentum for the RMF approximation.
If \(m_{\sigma}\) is \(500\rm \ MeV\), the maximum RMF-applicable density is around \(n_0\), which means the RMF approximation has already broken the analytic structure of EOM beyond \(n_0\) for neglecting the exchange symmetry, and even worse for \(\approx 4n_0\) here.

The reason why this analytic problem is not so obvious in past studies based on RMF approximation is that the multimeson couplings were usually introduced perturbatively, and therefore the solution of RMF EOM is close to the standard Green's function solution. In this way, even if the RMF approximation becomes not so good at high density, the solution of RMF EOM can still be found. However, in our approach, the multimeson couplings account for the hadron mass generation from spontaneous chiral symmetry breaking, thus they cannot be chosen to be so small, and this worsens the problem.

\begin{figure}[htpb]
    \centering
        \includegraphics[width=0.4\textwidth]{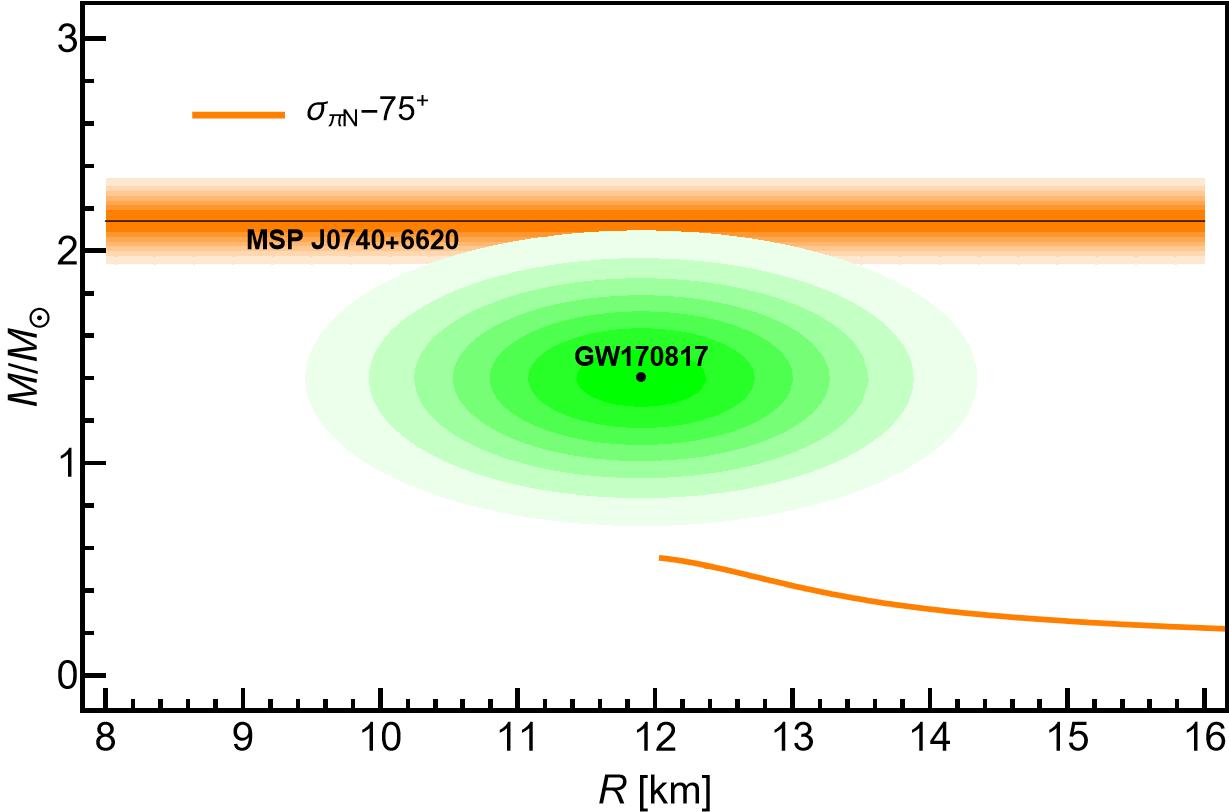}
        \hspace*{-0.8cm}\includegraphics[width=0.42\textwidth]{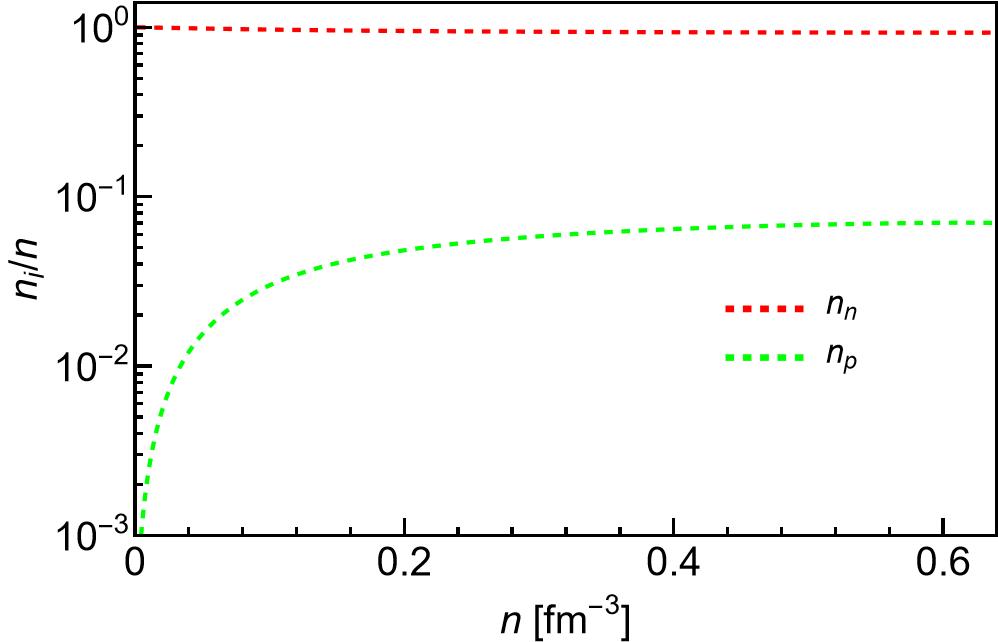}
    \caption{
        The NS structure with \(\sigma_{\pi N}\)-\(75^+\) parameter set. Upper: the M-R relation. Lower: the baryon fraction.
        The constraints of the NSs are from the MSP J0740+6620~\cite{NANOGrav:2019jur} and GW170817~\cite{LIGOScientific:2018cki}.
    }
    \label{fig:ns-p75}
\end{figure}
Moreover, when considering the $\beta$ equilibrium, it is found that hyperons do not emerge until \(4n_0\).
Using the EOS obtained with \(\sigma_{\pi N}\)-\(75^+\), when RMF approximation is applicable, the M-R relation of NSs can be calculated, and the results are shown in Fig.~\ref{fig:ns-p75}. It can be seen that the M-R relation from the solvable region of \(\sigma_{\pi N}\)-\(75^+\) cannot satisfy the astrophysical observations.

\begin{figure}[htpb]
    \centering
        \includegraphics[width=0.395\textwidth]{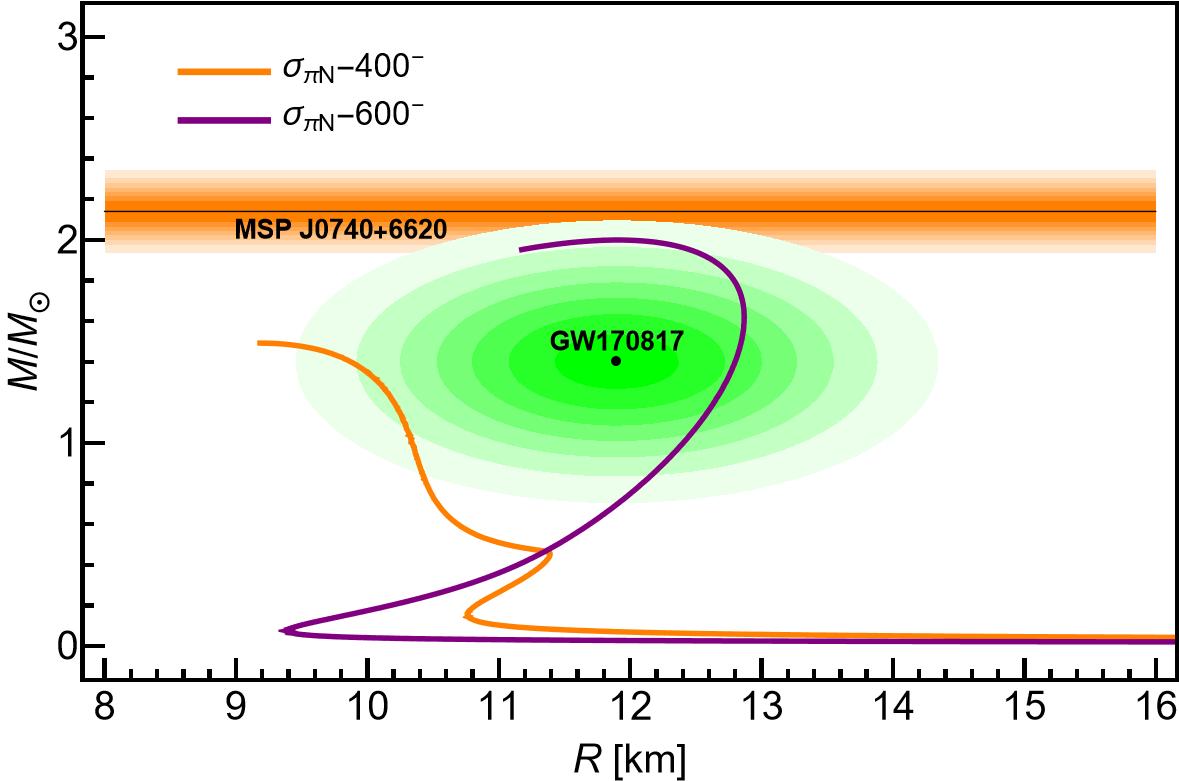}
        \hspace*{-0.75cm}\includegraphics[width=0.435\textwidth]{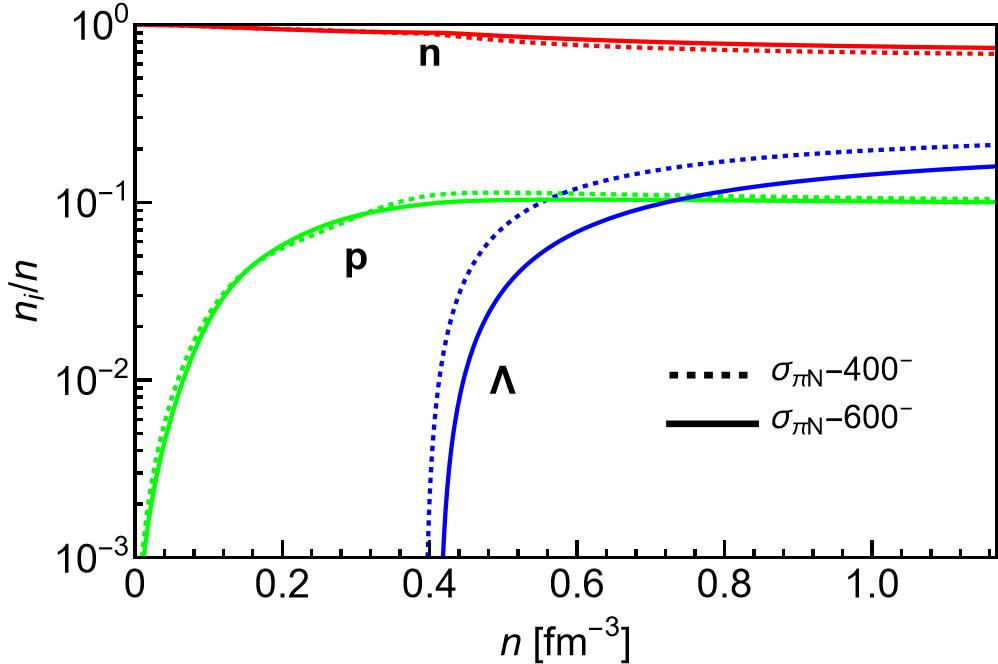}
    \caption{
        The NS structures with \(\sigma_{\pi N}\)-\(400^-\) and \(\sigma_{\pi N}\)-\(600^-\) parameter sets.
        The constraints are the same as in Fig.~\ref{fig:ns-p75}.
    }
    \label{fig:ns-neg}
\end{figure}

To find a reasonable M-R relation, we vary \(\sigma_{\pi N}\) by regarding it as a free parameter. It is found that the solution of the problem requires a negative value of \(\sigma_{\pi N}\), e.g., \(\sigma_{\pi N} \lesssim - 400\)~MeV. We present the mass spectra and NM properties for \(\sigma_{\pi N} \simeq - 100, -400\), and \(-600\)~MeV in Tables~\ref{tab:mass} and~\ref{tab:nm} and the NS structures for \(\sigma_{\pi N} \simeq -400\) and \(-600\)~MeV in Fig.~\ref{fig:ns-neg} because the solution problem still exists for \(\sigma_{\pi N} \simeq -100\)~MeV.
One can see that, with mass spectra and NM properties being well reproduced, the M-R relation gradually satisfies the astrophysical observations and the hyperons emerge at around \(2.5n_0\), with \(\sigma_{\pi N}\) becoming increasingly negative.
It is worth noting that such a requirement for a negative $\sigma_{\pi N}$ is specific to the present bELSM-$\xi$ framework where explicit chiral symmetry breaking terms are included in the three-flavor Lagrangian. In the two-flavor case studied in Ref.~\cite{Ma:2025llw}, no such solvable problem of the RMF EOM was found. This indicates that the explicit symmetry breaking terms in the three-flavor Lagrangian become significant at high densities, and the contribution of these terms to the baryon mass changes its value from the vacuum one.

Now, we understand that NS structure is sensitive to the value of \(\sigma_{\pi N}\) at densities. We next continue to improve the NS structure by adjusting the value of \(K(n_0)\) in the parameter sets as discussed in Ref.~\cite{Ma:2025llw}.
For this purpose, we take the incompressibility \(K(n_0)\) to be around \(500\rm \ MeV\) by tuning the magnitude of \(\tilde{g}_3\)~\cite{Ma:2025llw}, while fitting the mass spectra and NM properties [excluding \(K(n_0)\) from the fitting targets] as listed in Tables~\ref{tab:mass-K} and~\ref{tab:nm-K}.
It should be noted that such a large value of $K(n_0)\approx 500$~MeV, which exceeds the empirical range  constrained by giant monopole resonance experiments, is treated here as an exploratory parameter  rather than a physically preferred scenario. It reflects the tension between the physics at vacuum and low densities and that at high densities, which may stem from the limitations of the LO Lagrangian truncation and the RMF approach itself. The motivation for this exploration is that, as we will see in Sec.~\ref{sec:LamNS}, $K(n_0)$ plays an important role at high densities in determining the stiffness of the EOS, and an enlarged $K(n_0)$ is found to be effective in relieving the suppression of the maximum mass of NSs caused by the emergence of hyperons.
We find that the solution problem for \(\sigma_{\pi N}\)-\(100^-\) disappears but remains for \(\sigma_{\pi N}\)-\(75^+\), which addresses the behavior of explicit symmetry breaking at high densities: It should be negative at densities, at least in RMF approximation.
The mass spectra and NM properties produced in this case are listed in Tables~\ref{tab:mass-K} and~\ref{tab:nm-K}, respectively, with parameter sets listed in Table~\ref{tab:para-K}.
\begin{table}[htpb]
    \caption{
        The choices of parameters with \(K(n_0) \approx 500\)~MeV.
        The definitions of the symbols are the same as in Table~\ref{tab:para}.
        }
    \label{tab:para-K}
        \begin{tabular}{@{}ccccc}
    \hline
    \hline
    &\(\sigma_{\pi N}\)-\(100^{-}K\)&\(\sigma_{\pi N}\)-\(400^{-}K\)&\(\sigma_{\pi N}\)-\(600^{-}K\)\\
    \hline
    \(c_2(\times10^5\rm \ MeV^2)\)&$2.257$&$2.161$&$2.172$\\
    \hline
    \(G(\times10^4\rm \ MeV^2)\)&$2.623$&$2.623$&$2.623$\\
    \hline
    \(\alpha_3(\times10^{-2}\rm \ MeV)\)&$-7.602$&$-6.177$&$-5.376$\\
    \hline
    \(\alpha_8(\rm \ MeV)\)&$15.53$&$15.53$&$15.53$\\
    \hline
    \(\xi_0(\rm \ MeV)\)&$-26.10$&$53.13$&$132.5$\\
    \hline
    \(\tilde{h}_2\)&$245.7$&$356.0$&$438.7$\\
    \hline
    \(\tilde{g}_3\)&$34.88$&$115.8$&$198.4$\\
    \hline
    \(a_1\)&$302.5$&$-289$&$-466.7$\\
    \hline
    \(b_1\)&$-25.08$&$-20.55$&$-18.11$\\
    \hline
    \(b_2\)&$-145.1$&$-120.4$&$-106.0$\\
    \hline
    \(b_3\)&$113.3$&$-129.3$&$-262.4$\\
    \hline
    \(b_4\)&$-7.098$&$-5.861$&$-5.129$\\
    \hline
    \(b_5\)&$-0.2392$&$-0.2391$&$-0.2391$\\
    \hline
    \(g\)&$31.07$&$36.11$&$29.81$\\
    \hline
    \(g_{\sigma NN}\)&$-11.29$&$-14.51$&$-16.93$\\
    \hline
    \(g_{\omega NN}\)&$5.312$&$8.964$&$11.29$\\
    \hline
    \(g_{\rho NN}\)&$3.684$&$2.545$&$2.105$\\
    \hline
    \(x_{\omega}\)&$1.318$&$1.172$&$1.148$\\
    \hline
    \hline
    \end{tabular}
\end{table}

\begin{table*}[htpb]
    \caption{
        The mass spectra of relevant hadrons in the unit of \(\rm MeV\) from the parameter sets in Table~\ref{tab:para-K}.
        The definition of the symbols and the empirical values are the same as in Table~\ref{tab:mass}.
        }
    \label{tab:mass-K}
        \begin{tabular}{@{}cccccc}
    \hline
    \hline
    &Empirical&\(\sigma_{\pi N}\)-\(100^{-}K\)&\(\sigma_{\pi N}\)-\(400^{-}K\)&\(\sigma_{\pi N}\)-\(600^{-}K\)\\
    \hline
    \(m_n\)&$939.565\pm (5\times10^{-7})$&$939.565$&$939.565$&$939.565$\\
    \hline
    \(m_p\)&$938.272\pm (2.9\times10^{-7})$&$938.272$&$938.272$&$938.272$\\
    \hline
    \(m_{\Lambda}\)&$1115.68\pm(0.006)$&$1115.54$&$1115.57$&$1115.58$\\
    \hline
    \(m_{\Sigma^+}\)&$1189.37\pm0.06$&$1191.21$&$1191.74$&$1191.79$\\
    \hline
    \(m_{\Sigma^0}\)&$1192.64\pm0.02$&$1192.18$&$1192.84$&$1191.98$\\
    \hline
    \(m_{\Sigma^{-}}\)&$1197.45\pm0.03$&$1191.15$&$1191.95$&$1192.16$\\
    \hline
    \(m_{\Xi^{0}}\)&$1314.82\pm0.21$&$1330.65$&$1330.90$&$1330.90$\\
    \hline
    \(m_{\Xi^{-}}\)&$1321.70\pm0.09$&$1329.30$&$1329.82$&$1329.98$\\
    \hline
    \(m_{a_0}\)&$995.000\pm25.000$&$979.998$&$980.024$&$980.040$\\
    \hline
    \(m_{f_0}\)&$995.000\pm15.000$&$989.962$&$989.971$&$989.974$\\
    \hline
    \(m_{\sigma}\)&$475.000\pm75.000$&$984.993$&$986.010$&$985.019$\\
    \hline
    \(m_{\rho}\)&$773.000\pm2.000$&$775.260$&$775.260$&$775.260$\\
    \hline
    \(m_{\omega}\)&$782.660\pm0.13$&$782.660$&$782.660$&$782.660$\\
    \hline
    \(m_{\phi}\)&$1019.46\pm0.02$&$1019.46$&$1019.46$&$1019.46$\\
    \hline
    \(\sigma_{\pi N}\)&$60.5000\pm28.5000$&$-104.048$&$-402.087$&$-625.663$\\
    \hline
    \(m\)&$878.419\pm29.1465$&$1043.05$&$1341.09$&$1564.66$\\
    \hline
    \hline
    \end{tabular}
\end{table*}

\begin{table}[htpb]
    \caption{
        The properties of NM at saturation density $n_0$ from the parameter sets in Table~\ref{tab:para-K}.
        The conventions are the same as in Table~\ref{tab:nm}.
        }
    \label{tab:nm-K}
        \begin{tabular}{@{}cccccc}
    \hline
    \hline
    &Empirical&\(\sigma_{\pi N}\)-\(100^{-}K\)&\(\sigma_{\pi N}\)-\(400^{-}K\)&\(\sigma_{\pi N}\)-\(600^{-}K\)\\
    \hline
    \(n_0\)&$0.1550\pm0.0500$&$0.1592$&$0.1592$&$0.1592$\\
    \hline
    \(E_0\)&$-15.00\pm1.00$&$-16.00$&$-15.99$&$-16.00$\\
    \hline
    \(K\)&---&$521.9$&$520.3$&$518.8$\\
    \hline
    \(E_{\rm sym}\)&$30.90\pm1.90$&$30.00$&$30.00$&$30.01$\\
    \hline
    \(L\)&$52.50\pm17.50$&$80.15$&$93.28$&$76.64$\\
    \hline
    \hline
    \end{tabular}
\end{table}

From these tables, one can conclude that the mass spectra are barely affected by the variation of \(K(n_0)\), the properties of NM at saturation density are also reproduced, with only the symmetry energy slope \(L(n_0)\) becoming larger than the previous results, but still roughly within the empirical range.

\begin{figure}[htpb]
    \centering
        \includegraphics[width=0.395\textwidth]{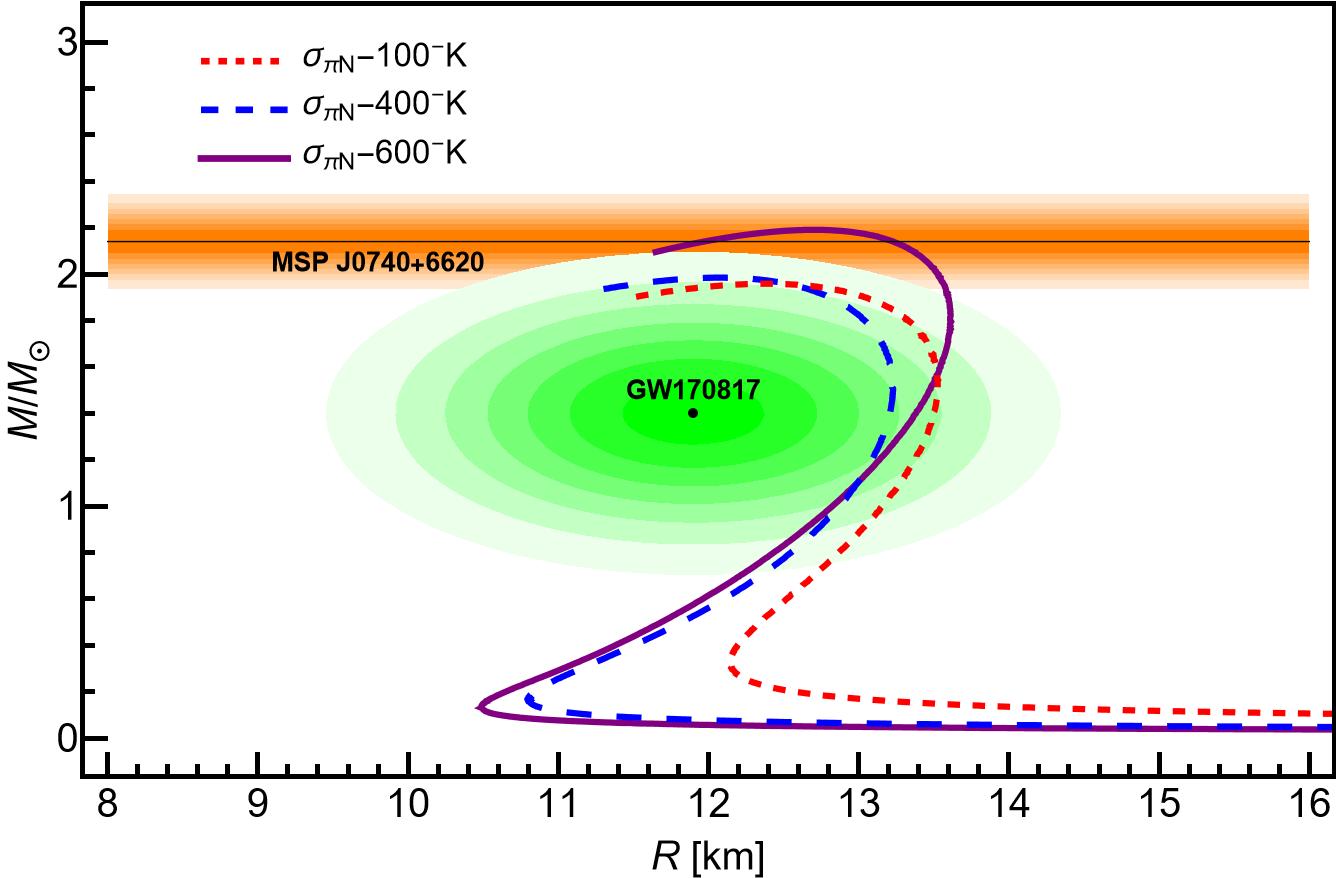}
        \hspace*{-0.60cm}\includegraphics[width=0.43\textwidth]{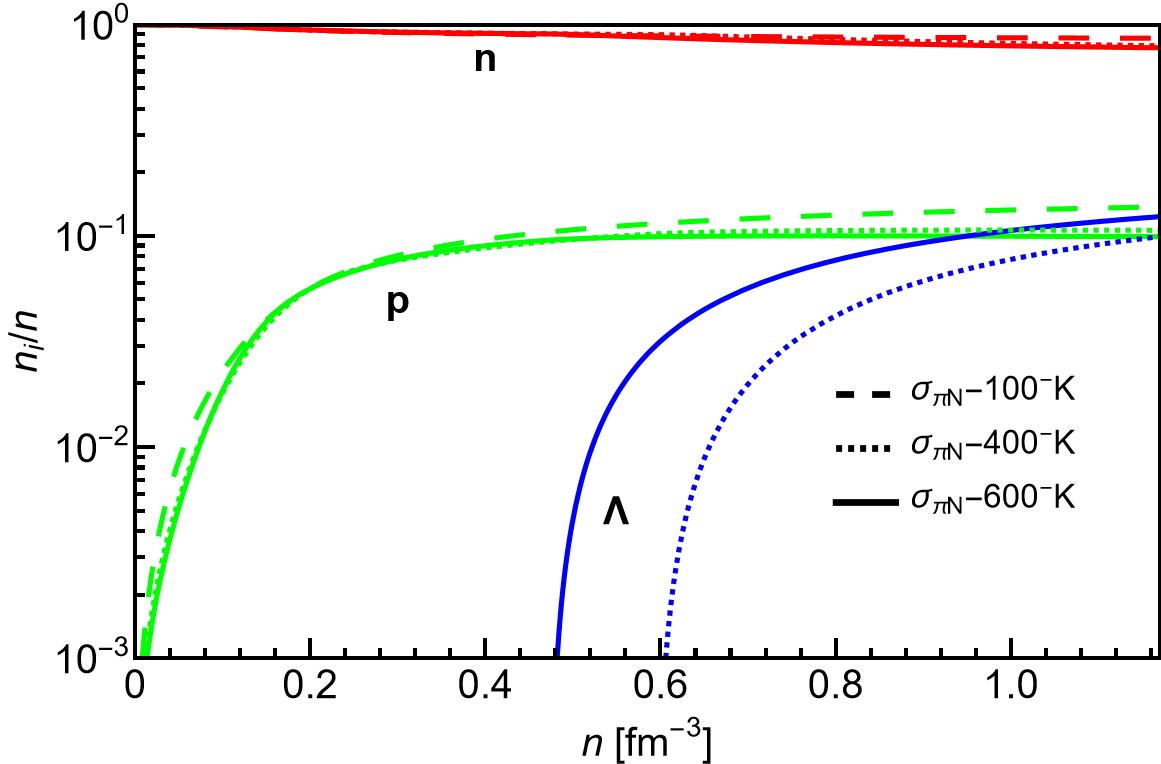}
    \caption{
        The NS structures with \(\sigma_{\pi N}\)-\(100^-K\), \(\sigma_{\pi N}\)-\(400^-K\), and \(\sigma_{\pi N}\)-\(600^-K\) parameter sets.
        The constraints are the same as in Fig.~\ref{fig:ns-p75}.
    }
    \label{fig:ns-K}
\end{figure}

The M-R relations and baryon compositions of NSs with the parameter sets in Table~\ref{tab:para-K} are shown in Fig.~\ref{fig:ns-K}. It can be seen that the M-R relations of all parameter sets with negative \(\sigma_{\pi N}\) are improved to satisfy the observations, and \(\sigma_{\pi N}\sim-600\rm \ MeV\) is still more favored by astrophysical data.
However, the baryon compositions of NM in Fig.~\ref{fig:ns-K} show that the hyperons will emerge above \(3n_0\), and, the more negative \(\sigma_{\pi N}\), the earlier emergence, in contrast to the previous cases with \(K(n_0)\sim 200\rm \ MeV\). However, in this case, for \(\sigma_{\pi N}\sim -100\rm \ MeV\), the hyperons still will not emerge at normal NS matter density.

\section{Effects of hyperons on neutron star structure}
\label{sec:LamNS}

After presenting the results of NS structures with three-flavor bELSM-\(\xi\), we compare the present results with the past studies of three-flavor NS matter to explore the implications of the current model.
Here, it should be clarified that the two-flavor case referred to in the comparisons below means the bELSM-\(\xi\) calculation of nuclear matter with hyperon emergence neglected, i.e., the same framework as the present work but without strange baryons. This is distinct from the previous bELSM studies in Refs.~\cite{Ma:2023eoz,Ma:2025llw}, which were constructed without the explicit symmetry breaking terms.
Here, we take the widely discussed Walecka-type model as the benchmark.
The details of the model and parameter choice are listed in Appendix~\ref{app:wal}. In the following, we make two comparisons.
First, we compare our results with the TM1 model with hyperon (TM1-H1), where the couplings of the \(\omega\) meson to hyperons and nucleons are almost identical, in order to see the difference between bELSM-\(\xi\) and Walecka-type models regardless of the breakings of SU(3) flavor symmetry. Second, the parameter set \(\sigma_{\pi N}\)-\(600^-K\), which gives the best NS structure in our framework, is compared with the Walecka-type models, TM1-H0.6, and parametrization $U_{\rm cut}(\sigma)$.

\begin{figure}[htpb]
    \centering
        \includegraphics[width=0.395\textwidth]{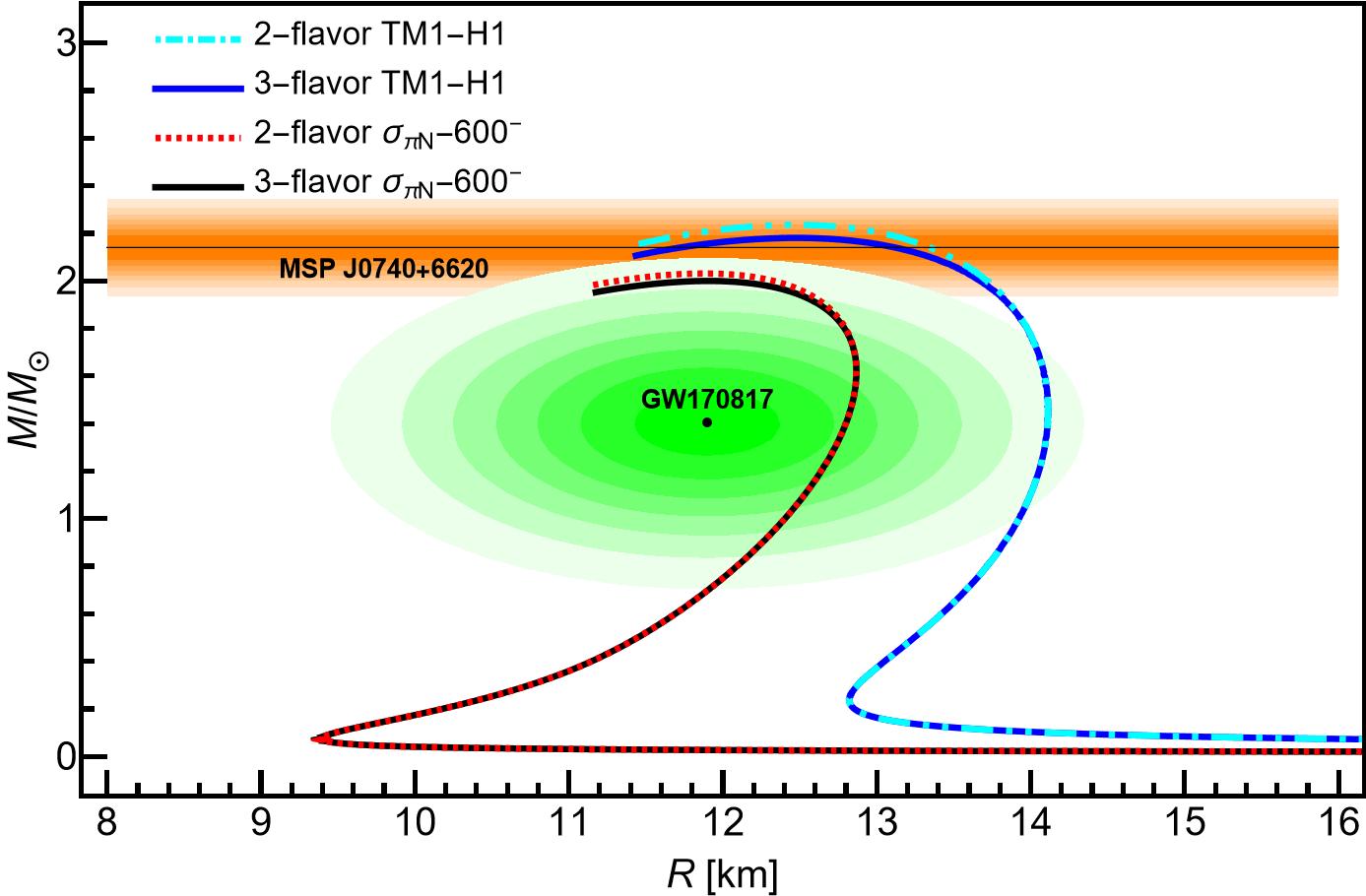}
        \hspace*{-0.60cm}\includegraphics[width=0.428\textwidth]{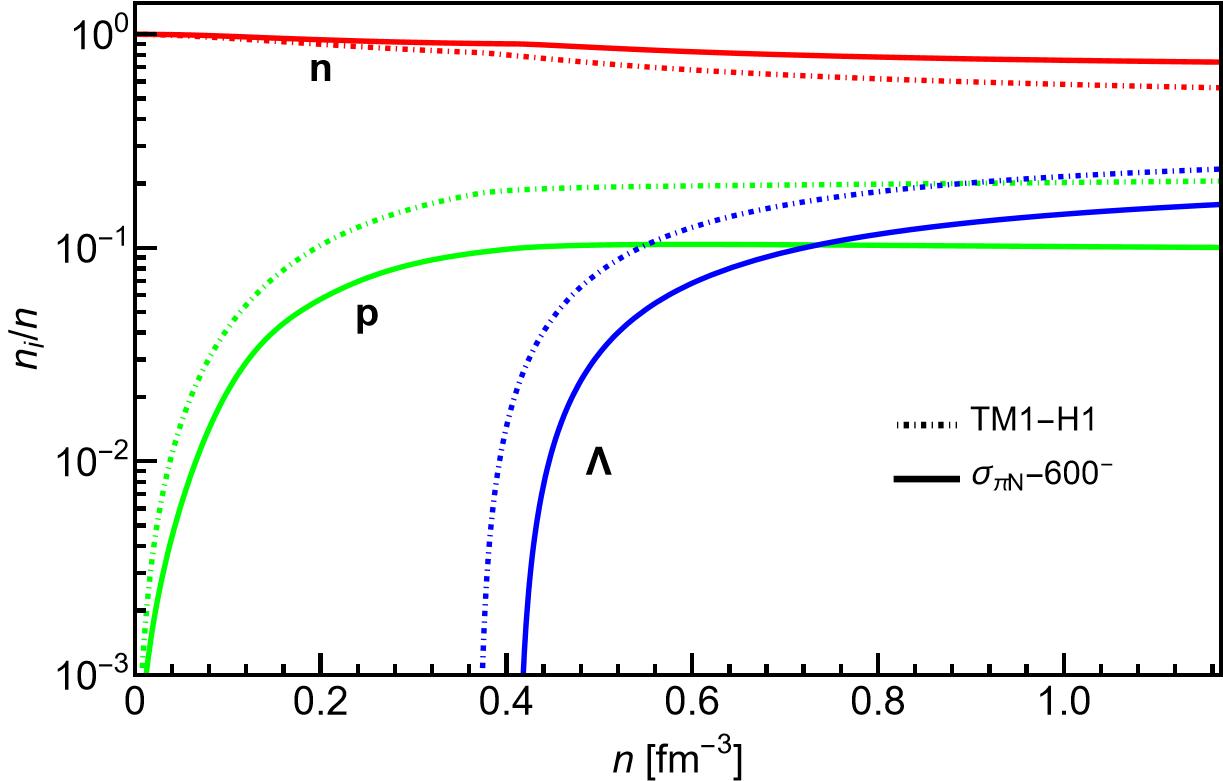}
    \caption{
        The NS structures with TM1-H1 and \(\sigma_{\pi N}\)-\(600^-\) parameter sets.
        The constraints are the same as in Fig.~\ref{fig:ns-p75}.
        Since \(x_{\sigma}= 1\) at LO bELSM-\(\xi\), TM1-H1 is chosen as the benchmark.
    }
    \label{fig:ns-wal}
\end{figure}

The first comparison of the NS structures is shown in Fig.~\ref{fig:ns-wal}. For consistency, the NM properties and the hyperon coupling \(g_{\sigma \Lambda\Lambda}/g_{\sigma NN}\) are constrained to be similar in two models. It can be seen that the M-R relation of \(\sigma_{\pi N}\)-\(600^-\) is the most compact one and favored by GW170817. The M-R relations of two- and three-flavor NSs in our framework are almost similar, except for the suppressed maximum mass of three-flavor NSs. The reason is that the hyperon emerges at a higher density in our framework, compared to the TM1-H1 set, and has fewer hyperons at high densities, resulting in less suppression on the maximum mass of NSs. In addition, in our framework, the maximum mass of \(\sigma_{\pi N}\)-\(600^-\) is slightly smaller than the central value of \(2M_{\odot}\) NS observations.

\begin{figure}[htpb]
    \centering
        \includegraphics[width=0.395\textwidth]{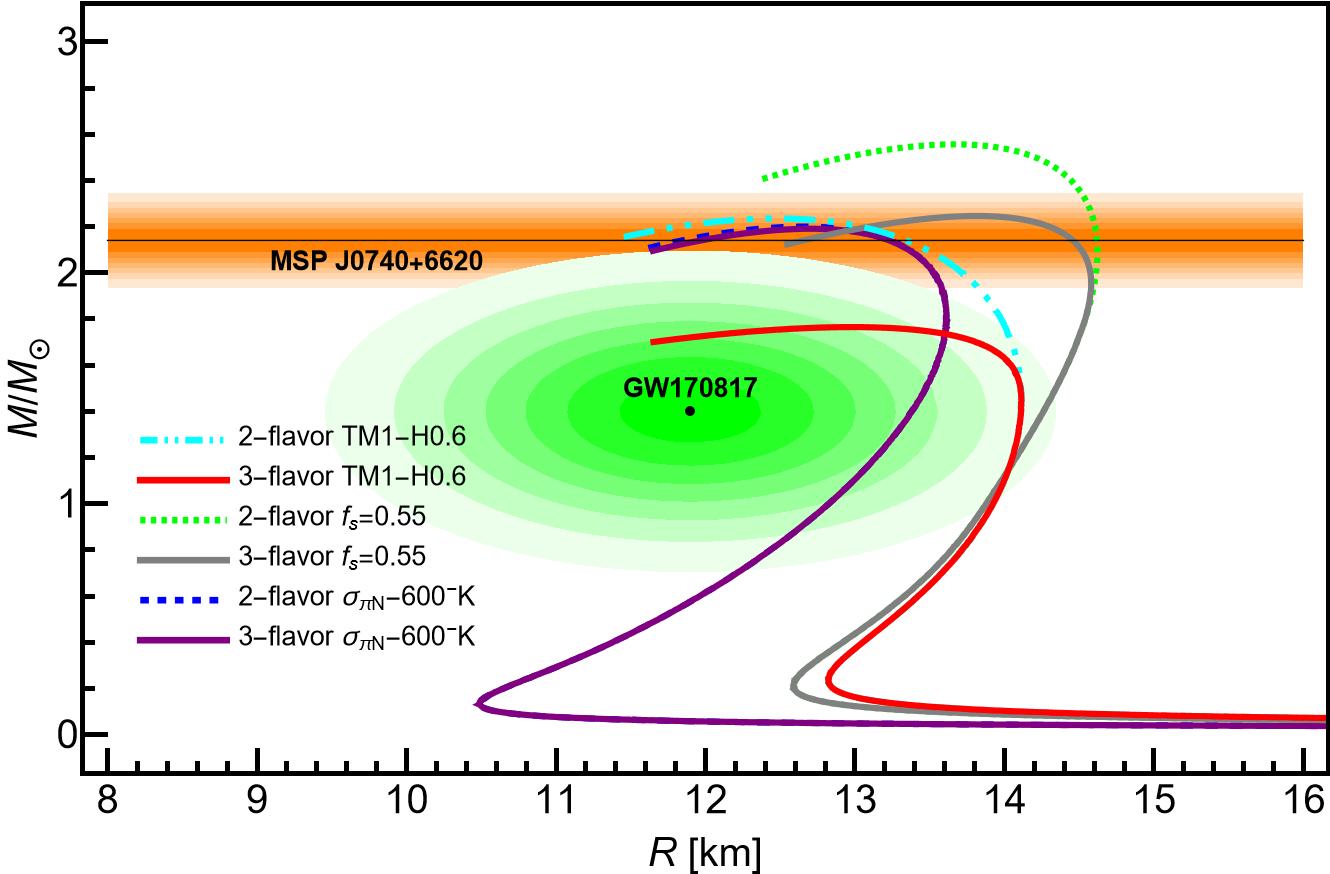}
        \hspace*{-0.15cm}\includegraphics[width=0.402\textwidth]{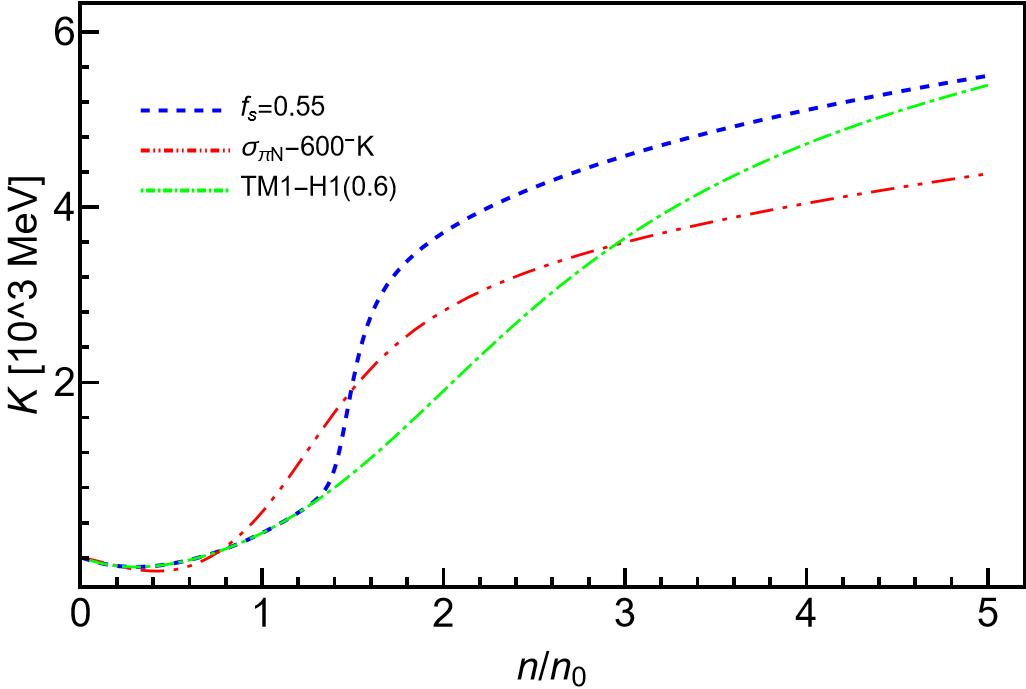}
    \caption{
        The NS structures with \(f_s=0.55\)~\cite{Zhang:2018lpl}, \(\sigma_{\pi N}\)-\(600^-K\), and TM1-H1(0.6) parameter sets of two- and three-flavor cases.
        The constraints are the same as in Fig.~\ref{fig:ns-p75}.
    }
    \label{fig:ns-cut}
\end{figure}

To heal the suppression of maximum mass when considering the hyperons in the core, incompressibility plays an important role (see, Figs.~\ref{fig:ns-neg} and \ref{fig:ns-K}). In the literature, a parametrization of scalar potential \(U_{\rm cut}(\sigma)\) was introduced to make EOS of NS matter stiffer at intermediate densities (around \(2n_0\)),~\cite{Zhang:2018lpl,Maslov:2015lma}.
The comparison of M-R relations between \(\sigma_{\pi N}\)-\(600^-K\), TM1-H0.6, and $U_{\rm cut}(\sigma)$ (the parameter set \(f_s=0.55\)~\cite{Zhang:2018lpl}) is presented in Fig.~\ref{fig:ns-cut}.
It can be seen that the M-R relation from \(\sigma_{\pi N}\)-\(600^-K\) is more compact than the one from both TM1-H0.6 and \(f_s=0.55\).

Now, let us look at both Figs.~\ref{fig:ns-wal} and~\ref{fig:ns-cut}. We can see that the maximum mass of a NS from \(\sigma_{\pi N}\)-\(600^-K\) is larger than that from \(\sigma_{\pi N}\)-\(600^-\). The \(U_{\rm cut}(\sigma)\), which can be reduced to the TM1 case at low densities but has larger incompressibility at high densities, also results in a larger maximum mass in the M-R relation, thus resolving the hyperon problem in TM1.
Then, we can understand that increasing \(K(n_0)\) in bELSM-\(\xi\) parameter sets has a similar effect to \(U_{\rm cut}(\sigma)\) in Walecka-type models, both make the EOS stiffer above $3n_0$, and therefore the maximum mass of NSs larger.
In addition, it is noticed that the emergence of hyperons does not reduce the maximum mass significantly in bELSM-\(\xi\), nor does it in TM1-H1.
However, for TM1-H0.6 and \(U_{\rm cut}(\sigma)\), the decreasing is obvious, which is due to \(g_{\sigma \Lambda\Lambda}/g_{\sigma NN}\approx0.6\).
However, these two couplings are identical at the LO of bELSM-\(\xi\), the same as TM1-H1.

\section{Summary and outlook}
\label{sec:conclusion}

In this work, we presented a new three-flavor baryon effective Lagrangian with explicit chiral symmetry breaking (termed as bELSM-\(\xi\)).
The framework is based on the chiral linear sigma model which has been applied to dense NM to investigate the properties of NM around \(n_0\) and the structure of NSs~\cite{Ma:2023eoz,Ma:2025llw}.
With LO bELSM-\(\xi\), the vacuum mass spectra of related baryons and mesons can be reproduced, and the properties of NM at saturation density can also be well described within RMF approximation.
However, when we tried to extend the EOS of NM beyond saturation density with the parameter set whose \(\sigma_{\pi N}\approx 75 \rm \ MeV\), we found that the RMF EOM have no real number solution above certain density, i.e., \(\simeq 4n_0\), and it is hard to give a reasonable M-R relation of NSs.
To solve this problem, we adjusted the value of \(\sigma_{\pi N}\) to be negative, and found that the M-R relation can be improved to satisfy the astrophysical observations and the hyperons emerge at about \(2.5n_0\).
\(\sigma_{\pi N}\) is suggested to be around \(-600\rm \ MeV\) to have a NS structure satisfying the astrophysical observations from GW170817~\cite{LIGOScientific:2018cki} and MSP J0740+6620~\cite{NANOGrav:2019jur}, at the LO calculation with the RMF approach.

Considering the importance of incompressibility \(K(n)\) in determining the M-R relation of NSs, 
we adjusted the value of \(K(n_0)\) to be around \(500\rm \ MeV\), and found that the M-R relation can be improved to describe the NSs around \(2M_{\odot}\) better with \(\sigma_{\pi N}\sim -100\rm \ MeV\), still negative.
It indicates that \(\sigma_{\pi N}\), labeling the contribution of explicit symmetry breaking to nucleon mass, should be negative at high densities, rather than \(75\rm \ MeV\), the empirical value at vacuum~\cite{Bernard:1995dp,Meissner:2005ba}, at the least from the LO bELSM-\(\xi\) with RMF.
After comparing to the past studies of three-flavor NS matter, we find that the M-R relations of bELSM-\(\xi\) are more compact and more favored by GW170817 constraints~\cite{LIGOScientific:2018cki} compared to the ones from Walecka-type models.
Increasing the value of \(K(n_0)\) in bELSM-\(\xi\) parameter sets leads to a larger maximum mass of NSs and describes the NSs around \(2M_{\odot}\) better, similar to the Walecka-type models with an additional \(U_{\rm cut}(\sigma)\) introduced~\cite{Zhang:2018lpl,Maslov:2015lma}.
Compared to both the Walecka-type RMF models and Brueckner-Hartree-Fock approaches, the bELSM-$\xi$ relates more closely to the underlying QCD symmetries through the chiral Lagrangian construction with explicit breaking included. However, the present results should be understood as the indication from the LO bELSM-$\xi$ with RMF, and the tension found here calls for a more systematic treatment beyond the LO in the future.

The comparison between M-R relations from two- and three-flavor models shows that the suppression of maximum mass after introducing hyperons is not significant in bELSM-\(\xi\), because \(g_{\sigma \Lambda\Lambda}=g_{\sigma NN}\) at LO bELSM-\(\xi\), making hyperons appear at a slightly higher density. These two couplings could be different in Walecka-type models~\cite{Schaffner:1995th,Zhang:2018lpl}.
The tension between the physics at vacuum/low densities and high densities was already found in the two-flavor case in Ref.~\cite{Ma:2025llw}. In the three-flavor case, this tension is sharpened by the explicit symmetry breaking terms: without the adjustment of $\sigma_{\pi N}$, the RMF EOM lose real number solutions above a certain density. The tuning of $\sigma_{\pi N}$ to negative values is therefore a parametric study indicating how the explicit breaking contribution to the baryon mass should evolve with density under the LO RMF treatment, rather than a direct prediction of the physical $\pi N$ sigma term at high densities.
The splitting of $g_{\sigma \Lambda\Lambda}$ and $g_{\sigma NN}$ can be included in next-to-leading order bELSM-\(\xi\) by introducing a term such as \({\rm Tr}(\bar{B}\Phi'\xi B)\).
In addition to the above concerns, the current framework also ignores the four-quark configurations of scalar mesons, which are crucial to the mass ordering of the \(\sigma\) meson~\cite{Fariborz:2005gm,Ma:2023eoz,ParticleDataGroup:2024cfk}.
In the future, we are going to systematically incorporate the next-to-leading order effect to have a more reasonable description of the three-flavor NS matter and investigate the relationship between microscopic symmetry patterns and macroscopic properties of NSs.

\section*{Acknowledgments}
The work of Y. M. is supported by Jiangsu Funding Program for Excellent Postdoctoral Talent under Grant No. 2025ZB516. Y.~L. M. is supported in part by the National Science Foundation of China (NSFC) under Grant No. 12547104, the National Key R\&D Program of China under Grant No. 2021YFC2202900, and Gusu Talent Innovation Program under Grant No. ZXL2024363.

\section*{Data availability}
There are no publicly available research data or software supporting this manuscript. Requests for further information or data should be sent to the authors.

\appendix

\section{The Lagrangian of bELSM-\(\xi\) at the leading order}\label{app:1st-L}

In this appendix, we give the full Lagrangian of the bELSM-\(\xi\). Following Ref.~\cite{Ma:2025llw}, we decompose the Lagrangian into three parts 
\be
\mathcal{L}=\mathcal{L}_M+\mathcal{L}_V+\mathcal{L}_B.
\ee

The scalar meson part \(\mathcal{L}_M\) is written as 
\be
\mathcal{L}_{\rm M} & = & \frac{1}{2}\operatorname{Tr}\left(\partial_\mu\Phi'\partial^\mu\Phi'^\dagger\right)+c_2 \operatorname{Tr}\left(\Phi'\Phi'^\dagger\right) \nonumber\\
& &{} -c_4 \operatorname{Tr}\left(\Phi'\Phi'^\dagger\Phi'\Phi'^\dagger\right) -\frac{b_1}{2} \operatorname{Tr}\left(\xi\Phi'^\dagger\Phi'\Phi'^\dagger+\rm h. c. \right) \nonumber\\
& &{} -G \operatorname{Tr}\left(\xi\Phi'^\dagger+\rm h. c.\right)-c_3\operatorname{ln}^2\left(\frac{\operatorname{det\Phi'}}{\operatorname{det\Phi'^\dagger}}\right)\ ,
\ee
where the \(c_3\) term accounts for the \(\rm U(1)_A\) breaking~\cite{Rosenzweig:1979ay} and the meson matrix  \(\Phi'\) is written in terms of scalar and pseudoscalar nonets as
\be
\Phi' & = & S'+iP \nonumber\\
& = & \frac{1}{\sqrt{2}}\sum_{i=1}^{8}(S'_i\lambda_i+iP_i\lambda_i)+\frac{1}{\sqrt{3}}(S'_0I+iP_0I) .
\ee

\begin{widetext}
The vector meson Lagrangian \(\mathcal{L}_V\), in terms of left-hand \(L^\mu = V_\mu - A_\mu\) and right-hand \(R^\mu = V_\mu + A_\mu\) fields, can be written as
\begin{eqnarray}
\mathcal{L}_V & = &{} -\frac{1}{2}\operatorname{Tr}\left(L_{\mu\nu}L^{\mu\nu}+R_{\mu\nu}R^{\mu\nu}\right)\nonumber\\
& &{} +g_1\operatorname{Tr}\left(\partial_\mu L_\nu L^\mu L^\nu+\partial_\mu R_\nu R^\mu R^\nu\right)+g_2\operatorname{Tr}\left(\partial_\mu L_\nu L^\nu L^\mu+\partial_\mu R_\nu R^\nu R^\mu\right)\nonumber\\
& & {} +g_3\operatorname{Tr}\left(L_\mu L_\nu L^\mu L^\nu+R_\mu R_\nu R^\mu R^\nu\right)+g_4\operatorname{Tr}\left(L_\mu L^\mu L_\nu L^\nu+R_\mu R^\mu R_\nu  R^\nu\right)\nonumber\\
& &{} +h_1\operatorname {Tr}\left(\Phi'\Phi'^\dagger L_\mu L^\mu+\Phi'^\dagger\Phi' R_\mu R^\mu\right)+h_2\operatorname{Tr}\left(L_\mu\Phi' R^\mu\Phi'^\dagger\right)+\frac{b'_2}{2}\operatorname {Tr}\left(\xi\Phi'^\dagger L_\mu L^\mu+\xi\Phi' R_\mu R^\mu\right)\nonumber\\
& & {} +\frac{\bar{b}_2}{2}\operatorname {Tr}\left(\Phi'\xi L_\mu L^\mu+\Phi'^\dagger\xi R_\mu R^\mu\right)+\tilde{b}_2\operatorname{Tr}\left(L_\mu\xi R^\mu\Phi'^\dagger\right)+\hat{b}_2\operatorname{Tr}\left(L_\mu\Phi' R^\mu\xi\right)\nonumber\\
& &{} +\frac{a_1}{2}\epsilon^{ijk}\epsilon^{lmn}\left[\left(L_\mu\right)_{il}\left(L^\mu\right)_{jm}\left(\Phi'\Phi'^\dagger\right)_{kn}+\left(R_\mu\right)_{il}\left(R^\mu\right)_{jm}\left(\Phi'^\dagger\Phi'\right)_{kn}\right]\nonumber\\
& &{} +\frac{a_2}{2}\epsilon^{ijk}\epsilon^{lmn}\left[\left(L_\mu\right)_{il}\left(L_\nu\right)_{jm}\left(L^\mu L^\nu\right)_{kn}+\left(R_\mu\right)_{il}\left(R_\nu\right)_{jm}\left(R^\mu R^\nu\right)_{kn}\right]\nonumber\\
& &{} +\frac{a_3}{2}\epsilon^{ijk}\epsilon^{lmn}\left[\left(L_\mu\right)_{il}\left(L^\mu\right)_{jm}\left(L_\nu L^\nu\right)_{kn}+\left(R_\mu\right)_{il}\left(R^\mu\right)_{jm}\left(R_\nu R^\nu\right)_{kn}\right]\nonumber\\
& &{} +\frac{a_4}{2}\epsilon^{ijk}\epsilon^{lmn}\left[\left(L_\mu\right)_{il}\left(L_\nu\right)_{jm}\left(\partial^\mu L^\nu\right)_{kn}+\left(R_\mu\right)_{il}\left(R_\nu\right)_{jm}\left(\partial^\mu R^\nu\right)_{kn}\right]\nonumber\\    
& &{} +b'_3\epsilon^{ijk}\epsilon^{lmn}\left[\left(L_\mu\right)_{il}\left(L^\mu\right)_{jm}\left(\xi\Phi'^\dagger\right)_{kn}+\left(R_\mu\right)_{il}\left(R^\mu\right)_{jm}\left(\xi\Phi'\right)_{kn}\right]\nonumber\\  
& &{} +\tilde{b}_3\epsilon^{ijk}\epsilon^{lmn}\left[\left(L_\mu\right)_{il}\left(L^\mu\right)_{jm}\left(\Phi'\xi\right)_{kn}+\left(R_\mu\right)_{il}\left(R^\mu\right)_{jm}\left(\Phi'^\dagger\xi\right)_{kn}\right]\ ,
\end{eqnarray}
where $L_{\mu\nu}=\partial_\mu L_\nu-\partial_\nu L_\mu$ and $R_{\mu\nu}=\partial_\mu R_\nu-\partial_\nu R_\mu$.

The baryon Lagrangian \(\mathcal{L}_B\) under the quark-diquark scheme is as follows:
\be
\mathcal{L}_B & = & \operatorname{Tr}\left(\bar{B}i\gamma_\mu\partial^\mu B\right) + c\operatorname{Tr}\left(\bar{B}\gamma_\mu V^\mu B+\bar{B}\gamma_\mu\gamma_5 A^\mu B\right)+c'\operatorname{Tr}\left(\bar{B}\gamma_\mu BV^\mu\right)+c'_A\operatorname{Tr}\left(\bar{B}\gamma_\mu\gamma_5 BA^\mu\right)\nonumber\\
& &{} + h\epsilon^{ijk}\epsilon^{lmn}\left[\left(\bar{B}\right)_{il}\gamma_\mu\left(B\right)_{jm}\left(V^\mu\right)_{kn}+\left(\bar{B}\right)_{il}\gamma_\mu\gamma_5\left(B\right)_{jm}\left(A^\mu\right)_{kn}\right] \nonumber\\
& &{} - \frac{g}{2}\operatorname{Tr}\left[\bar{B}\left(\Phi'+\Phi'^\dagger\right)B+\bar{B}\gamma_5\left(\Phi'-\Phi'^\dagger\right)B\right]\nonumber\\
& &{} -\frac{e}{2}\epsilon^{ijk}\epsilon^{lmn}\left[\left(\bar{B}\right)_{il}\left(\Phi'+\Phi'^\dagger\right)_{jm}\left(B\right)_{kn}-\left(\bar{B}\right)_{il}\gamma_5\left(\Phi'-\Phi'^\dagger\right)_{jm}\left(B\right)_{kn}\right]\nonumber\\
& &{} -b_4\operatorname{Tr}\left(\bar{B}\xi B\right)-b_5\epsilon^{ijk}\epsilon^{lmn}\left[\left(\bar{B}\right)_{il}\left(\xi\right)_{jm}\left(B\right)_{kn}\right]\ .
\ee
\end{widetext}

The relationship between the coefficients of the RMF Lagrangian used in this work and the first order bELSM-\(\xi\) Lagrangian is obtained as follows:
\be
\tilde{h}_2 & = & 2h_1+h_2,\ \tilde{g}_3 = 2\left(g_3+g_4\right),\nonumber\\ b_2 & = & b'_2+\bar{b}_2+\tilde{b}_2+\hat{b}_2, b_3 = 2\left(b'_3+\tilde{b}_3\right),\nonumber\\ \tilde{a}_2 & = & a_2+a_3\ .
\ee

\section{EOS of neutron star matter under RMF approximation}
\label{app:eos}

In addition to the hadronic Lagrangian introduced in Sec.~\ref{sec:model}, electrons should be introduced to account for the possible $\beta$ equilibrium inside the NSs.
The free Lagrangian of electrons is
\begin{equation}
    \mathcal{L}_e=\bar{e}\left(i\slashed{\partial}-m_e\right)e\ .
\end{equation}

Considering that vector meson \(\phi\) is heavier than \(1~\rm GeV\), we will not consider its contribution. Then,
under the RMF approximation, the EOM for the mesonic fields are
\begin{widetext}
\be
\label{eq:eom-rho}
{\rm EOM}_{\rho}(\sigma,\omega,\rho) & = & m_\rho^2 \rho+\frac{\tilde{g}_3}{2} \rho^3+\frac{3 \tilde{g}_3}{2} \rho \omega^2+\frac{\tilde{h}_2-a_1}{3} \rho \sigma^2+\frac{2 \sqrt{2} \tilde{h}_2\alpha_3+\sqrt{2} b_2 \xi_3}{2 \sqrt{3}} \omega \sigma \nonumber\\
& &{} + \frac{2 \sqrt{2}\left(\tilde{h}_2+2 a_1\right)\alpha_8+4\left(\tilde{h}_2-a_1\right)\alpha+\sqrt{2}\left(b_2+2 b_3\right) \xi_8+2\left(b_2-b_3\right) \xi_0}{6} \rho \sigma \nonumber\\
& &{} + \frac{\left[2 \tilde{h}_2\alpha_8+2 \sqrt{2} \tilde{h}_2\alpha+\sqrt{2} b_2 \xi_0+b_2 \xi_8\right]\alpha_3+b_2\left[\alpha_8+\sqrt{2}\alpha\right] \xi_3}{2 \sqrt{3}} \omega \nonumber\\
& &{} - g_{\rho N N}\left(n_p-n_n\right)=0\ ,
\ee
\be
\label{eq:eom-omega}
{\rm EOM}_{\omega}(\sigma,\omega,\rho) & = & m_\omega^2 \omega +\frac{\tilde{g}_3}{2} \omega^3+\frac{3 \tilde{g}_3}{2} \rho^2 \omega+\frac{\tilde{h}_2+a_1}{3} \omega \sigma^2+\frac{2 \sqrt{6} \tilde{h}_2\alpha_3+\sqrt{6} b_2 \xi_3}{6} \rho \sigma \nonumber\\
& &{} +\frac{2 \sqrt{2}\left(\tilde{h}_2-2 a_1\right)\alpha_8+2\left(\tilde{h}_2+a_1\right)\alpha+\sqrt{2}\left(b_2-2 b_3\right) \xi_8+2\left(b_2+b_3\right) \xi_0}{6} \omega \sigma \nonumber\\
& &{} +\frac{\left[2 \tilde{h}_2\alpha_8+2 \sqrt{2} \tilde{h}_2\alpha+\sqrt{2} b_2 \xi_0+b_2 \xi_8\right]\alpha_3+b_2\left[\alpha_8+\sqrt{2}\alpha\right] \xi_3}{2 \sqrt{3}} \rho \nonumber\\
& &{} -g_{\omega NN}\left(n_p+n_n\right)-g_{\omega \Lambda \Lambda} n_{\Lambda}=0\ ,
\ee
\be
\label{eq:eom-sigma}
{\rm EOM}_{\sigma}(\sigma,\omega,\rho) & = &{} - m_\sigma^2 \sigma-\frac{4 c_4}{3} \sigma^3+\frac{\tilde{h}_2-a_1}{3} \rho^2 \sigma+\frac{\tilde{h}_2+a_1}{3} \omega^2 \sigma \nonumber\\
& &{} -\left[4 c_4\alpha+b_1 \xi_0\right] \sigma^2+\frac{2 \sqrt{6} \tilde{h}_2\alpha_3+\sqrt{6} b_2 \xi_3}{6} \rho \omega \nonumber\\
& &{} +\frac{2 \sqrt{2}\left(\tilde{h}_2+2 a_1\right)\alpha_8+4\left(\tilde{h}_2-a_1\right)\alpha+\sqrt{2}\left(b_2+2 b_3\right) \xi_8+2\left(b_2-b_3\right) \xi_0}{12} \rho^2 \nonumber\\
& &{} +\frac{2 \sqrt{2}\left(\tilde{h}_2-2 a_1\right)\alpha_8+2\left(\tilde{h}_2+a_1\right)\alpha+\sqrt{2}\left(b_2-2 b_3\right) \xi_8+2\left(b_2+b_3\right) \xi_0}{12} \omega^2 \nonumber\\
& &{} +g_{\sigma N N} \sum_{i=p, n} \frac{\left(m_i-g_{\sigma N N} \sigma\right)^3}{\pi^2} \int_0^{\frac{k_i}{m_i-g_{\sigma N N} \sigma}} \frac{x^2}{\sqrt{1+x^2}} {\rm d} x \nonumber\\
& &{} +g_{\sigma \Lambda\Lambda} \frac{\left(m_\Lambda-g_{\sigma \Lambda\Lambda} \sigma\right)^3}{\pi^2} \int_0^{\frac{k_\Lambda}{m_\Lambda-g_{\sigma \Lambda\Lambda} \sigma}} \frac{x^2}{\sqrt{1+x^2}} {\rm d} x =0\ ,
\ee
where \(m_\sigma\) , \(m_\omega\), and \(m_\rho\) are calculated after spontaneous symmetry breaking, see Eq.~\eqref{eq:vev}, via
\begin{equation}
    m_{\sigma}^2=-\left.\frac{\partial^2\mathcal{L}_{\rm M}}{\partial \sigma\partial\sigma}\right|_{\sigma=\omega=\rho=0}, \quad m_{\omega}^2=\left.\frac{\partial^2\mathcal{L}_{\rm V}}{\partial \omega\partial\omega}\right|_{\sigma=\omega=\rho=0}, \quad m_{\rho}^2=\left.\frac{\partial^2\mathcal{L}_{\rm V}}{\partial \rho\partial\rho}\right|_{\sigma=\omega=\rho=0}\ .
\end{equation}
The meson fields can be solved by the above EOM with Fermion number densities \(n_{n}\), \(n_p\), and \(n_{\Lambda}\) input. After replacing the meson fields with these solutions, the EOM of baryons can be solved~\cite{Serot:1984ey} with free EOM for electrons. Then, after subtracting the vacuum constant Hamiltonian density, the Hamiltonian density of the system can be obtained as
\begin{eqnarray}
\mathcal{H} & = & \sum_{i=p, n} \frac{\left(m_i-g_{\sigma N N} \sigma\right)^4}{\pi^2} \int_0^{\frac{k_i}{m_i-g_{\sigma N N} \sigma}} x^2 \sqrt{1+x^2} d x \nonumber\\
& &{} +\frac{\left(m_\Lambda-g_{\sigma \Lambda\Lambda} \sigma\right)^4}{\pi^2} \int_0^{\frac{k_\Lambda}{m_\Lambda-g_{\sigma \Lambda\Lambda} \sigma}} x^2 \sqrt{1+x^2} d x + \frac{m_e^4}{\pi^2} \int_0^{\frac{k_e}{m_e}} x^2 \sqrt{1+x^2} d x \nonumber \\
& &{} + \frac{1}{2} m_\rho^2 \rho^2 + \frac{1}{2} m_\omega^2 \omega^2 + \frac{1}{2} m_\sigma^2 \sigma^2 + \frac{4 c_4\alpha_0 + b_1 \xi_0}{3} \sigma^3 + \frac{c_4}{3} \sigma^4 + \frac{3 \tilde{g}_3}{8} \rho^4 + \frac{3 \tilde{g}_3}{8} \omega^4 \nonumber \\
& &{} + \frac{\sqrt{3}\left[2 \tilde{h}_2\alpha_8 + 2 \sqrt{2} \tilde{h}_2\alpha_0 + b_2 \xi_8 + \sqrt{2} b_2 \xi_0\right]\alpha_3 + \sqrt{3}\left[b_2\alpha_8 + \sqrt{2} b_2\alpha_0\right] \xi_3}{6} \rho \omega \nonumber \\
& &{} + \frac{2 \sqrt{2}\left(\tilde{h}_2 + 2 a_1\right)\alpha_8 + 4\left(\tilde{h}_2 - a_1\right)\alpha_0 + \sqrt{2}\left(b_2 + 2 b_3\right) \xi_8 + 2\left(b_2 - b_3\right) \xi_0}{12} \rho^2 \sigma \nonumber \\
& &{} + \frac{2 \sqrt{2}\left(\tilde{h}_2 - 2 a_1\right)\alpha_8 + 2\left(\tilde{h}_2 + a_1\right)\alpha_0 + \sqrt{2}\left(b_2 - 2 b_3\right) \xi_8 + 2\left(b_2 + b_3\right) \xi_0}{12} \omega^2 \sigma \nonumber \\
& &{} + \frac{2 \sqrt{6} \tilde{h}_2\alpha_3 + \sqrt{6} b_2 \xi_3}{6} \rho \omega \sigma + \frac{\tilde{h}_2 - a_1}{6} \rho^2 \sigma^2 + \frac{\tilde{h}_2 + a_1}{6} \omega^2 \sigma^2 + \frac{9 \tilde{g}_3}{4} \rho^2 \omega^2\ .
\label{eq:H}
\end{eqnarray}

The $\beta$ equilibrium of the system are guaranteed by following relations:
\begin{equation}
    \mu_n=\mu_p+\mu_e,\quad \mu_\Lambda=\mu_n,\quad n_p=n_e\ ,
\end{equation}
where \(\mu_i\) is the chemical potential of Fermion \(i\), defined as \(\mu_i=\frac{\partial\mathcal{H}}{\partial n_i}\).

\section{EOS from Walecka-type models}
\label{app:wal}

The Walecka-type model considered is the following~\cite{Dutra:2014qga}:
\be
\mathcal{L}_{\rm W} & = & \bar{N}\left[i \gamma_\mu \partial^\mu-m_N+g_{\sigma N N} \sigma-g_{\omega_{N N}} \gamma_\mu \omega^\mu-g_{\rho_{N N}} \gamma_\mu \rho^{\mu a} \tau^a\right] N \nonumber\\
& &{} +\bar{\Lambda}\left[i \gamma_\mu \partial^\mu-m_{\Lambda}+g_{\sigma \Lambda \Lambda} \sigma-g_{\omega \Lambda \Lambda} \gamma_\mu \omega^\mu\right] \Lambda \nonumber\\
& &{} +\frac{1}{2}\left(\partial_\mu \sigma \partial^\mu \sigma-m_\sigma^2 \sigma^2\right)-\frac{1}{3} A \sigma^3-\frac{1}{4} B \sigma^4 \nonumber\\
& &{} -\frac{1}{4} F_{\mu\nu}F^{\mu\nu}+\frac{1}{2} m_\omega^2 \omega_\mu \omega^\mu+\frac{1}{4} c_3\left(\omega_\mu \omega^\mu\right)^2 - \frac{1}{4}B_{\mu\nu}^a B^{\mu\nu a}+\frac{1}{2} m_\rho^2 \rho^{\mu a} \rho_\mu^a\ .
\label{eq:LagWalecka}
\ee
The parameters in this model are chosen to reproduce the NM properties at saturation point and hyperon potential \(U_{\Lambda}=-28\rm \ MeV\) based on the TM1 set~\cite{Sugahara:1993wz}.
The parameter sets are listed in Table~\ref{tab:para-wal}, and NM properties are listed in Table~\ref{tab:nm-wal}.
\begin{table*}[htbp]
    \caption{
        The choices of the parameters in model~\eqref{eq:LagWalecka}.
        The masses are in units of \(\rm MeV\).
        \(A\) is in units of \(\rm fm^{-1}\), and the other couplings are dimensionless.
        \(x_{\sigma}=g_{\sigma\Lambda\Lambda}/g_{\sigma NN}\).
        \(x_{\sigma}=1.000(0.6000)\) for TM1-H1(0.6), respectively.
        }
    \label{tab:para-wal}
        \begin{tabular}{@{}ccccccccccccc}
    \hline
    \hline
    &\(m_N\)&\(m_{\Lambda}\)&\(m_{\omega}\)&\(m_{\rho}\)&\(m_{\sigma}\)&\(A\)&\(B\)&\(c_3\)&\(g_{\rho NN}\)&\(g_{\omega NN}\)&\(g_{\sigma NN}\)&\(x_{\omega}\)\\
    \hline
    TM1-H1(0.6)&$938.0$&$1116$&$783.0$&$770.0$&$511.2$&$-7.233$&$0.6183$&$71.31$&$4.632$&$12.61$&$-10.03$&\(1.149(0.6477)\)\\
    \hline
    \hline
    \end{tabular}
\end{table*}

\begin{table*}[htpb]
    \caption{
        The properties of NM at saturation density \(n_0\) from the parameters in Table~\ref{tab:para-wal}.
        The conventions are the same as in Table~\ref{tab:nm}.
        TM1-H1 and TM1-H0.6 only have different values of \(g_{\sigma \Lambda\Lambda}\), which does not affect the NM properties at saturation density.
        }
    \label{tab:nm-wal}
        \begin{tabular}{@{}cccccc}
    \hline
    \hline
    &\(n_0\)&\(e_0\)&\(K\)&\(E_{\rm sym}\)&\(L(n_0)\)\\
    \hline
    Empirical&$0.1550\pm0.0500$&\(-15.00\pm1.00\)&\(230.0\pm30.0\)&$31.90\pm1.90$&$52.50\pm17.50$\\
    \hline
    TM1-H1(0.6)&$0.1452$&\(-16.26\)&$281.2$&$31.89$&$110.8$\\
    \hline
    \hline
    \end{tabular}
\end{table*}

\end{widetext}

\bibliography{draft}

\end{document}